\documentclass[aps,prb,twocolumn,notitlepage]{revtex4-2}
\pdfoutput=1 
\usepackage{graphicx}
\usepackage[justification=raggedright]{caption}
\usepackage{xcolor}
\usepackage{enumerate}
\usepackage{amsmath, amssymb}
\usepackage{stmaryrd}
\usepackage{multirow}
\usepackage{float}
\usepackage{subcaption}
\usepackage{dsfont}
\usepackage{placeins}
\usepackage{hyperref}
\hypersetup{colorlinks,colorlinks=true,linkcolor=blue,citecolor=blue}

\DeclareMathAlphabet\mathbfcal{OMS}{cmsy}{b}{n}

\begin{document}

\title{Orbital magnetic quadrupole moment in higher order topological phases}

\author{Jacopo Gliozzi, Mao Lin, and Taylor L. Hughes}
\affiliation{Department of Physics and Institute for Condensed Matter Theory, University of Illinois at Urbana-Champaign, IL 61801, USA}
\begin{abstract}
We study the orbital magnetic quadrupole moment (MQM) in three dimensional higher-order topological phases. Much like electric quadrupole moment, which is associated with a charge response on the boundaries of a finite sample, the diagonal components of the MQM manifest as surface-localized magnetization and hinge currents. The hinge current is generally not equal to the difference of surface magnetizations that intersect at the hinge, and we show this mismatch is precisely quantified by the bulk MQM. We derive a quantum mechanical formula for the layer-resolved magnetization in slab geometries and use it to define the MQM of systems with gapped boundaries. Our formalism is then applied to several higher-order topological phases, and we show that the MQM can distinguish phases in some intrinsic and boundary-obstructed higher-order topological insulators. We then show that derivatives of the MQM with respect to the chemical potential can act as quantized topological invariants, similar to obtaining the 2D Chern number as a derivative of the magnetization with respect to the chemical potential. These invariants provide a new way to characterize 3D time-reversal breaking insulators that have vanishing magnetization.
\end{abstract}

\maketitle

\tableofcontents

\section{Introduction}

Macroscopic multipole moments, both electric and magnetic, are among the most fundamental properties of material{s}~\cite{raabbook}. In classical electromagnetism, they characterize the inhomogeneous distribution of charges and currents, upon which the phenomenological descriptions of, among others, ferroelectrics and ferromagnets are based~\cite{landau1984vol}. 	
Despite its seemingly simple nature, however, a precise understanding of the electric polarization of extended quantum solids remained elusive until the early 1990s~\cite{king1993theory,vanderbilt1993electric}. 
We now understand that the electric polarization can be determined as a geometric Berry phase of the electronic wavefunctions in the momentum space Brillouin zone (BZ). This approach has been used to predict various dielectric responses in real materials~\cite{resta1994macroscopic}, such as piezoelectricity in GaAs~\cite{vanderbilt1993electric} and spontaneous polarization in ferroelectric
perovskites~\cite{resta1993towards}, and is also closely related to the physics of topological insulator{s}~\cite{hasan2010colloquium,qi2011topological}. For example, symmetries that quantize the electric polarization~\cite{zak1989} can protect topological crystalline insulator phases~\cite{hughes2011inversion,turner2012}. Additionally, the change in polarization after an adiabatic cycle describes quantized charge pumping characterized by the Chern number~\cite{thouless1983quantization}. 

{The topological nature of such quantized charge transport is closely related to the fact that electric dipole moment is a geometric phase quantity determined by the zero-temperature ground state. } Magnetic dipole {moment}, or magnetization, on the other hand, is thermodynamic {in} nature~\cite{ashcroft1976solid}. Indeed, polarization, as a bulk material property, only makes sense in insulators, which strictly speaking only exist at zero temperature in the absence of disorder or interactions~\cite{resta2010electrical}.  In contrast, many more materials, including {Mott} insulators and metals, have well-defined magnetization at zero or finite temperature~\cite{thonhauser2011theory}.
Both the spin and the orbital motion of electrons can contribute to the total magnetization, and while the unit-cell average of the spin contribution is well-defined in the thermodynamic limit~\cite{resta2010electrical,aryasetiawan2019modern}, this is not the case for the orbital contribution. Specifically, the presence of the unbounded position operator in its definition renders it ill-defined in {extended} systems. As such, a microscopic calculation of magnetization is a nontrivial task in bulk materials. 

Fortunately, this problem was eventually resolved using the Wannier-function {re}presentation of an  insulator~\cite{thonhauser2005orbital} and using a semiclassical approach for Bloch electrons~\cite{xiao2005berry}. These approaches {were} later generalized to finite temperature~\cite{xiao2006berry} for both insulators and metals {with multiple bands}~\cite{Ceresoli2006}. A collective effort eventually {led} to a fully quantum mechanical treatment of the orbital magnetization~\cite{shi2007quantum}, which can be applied to insulators (having any Chern number) and metal{s} at arbitrary temperature. A Green function approach to calculate orbital magnetization has also been put forward~\cite{nourafkan2014orbital}, and later generalized to the many-body case~\cite{aryasetiawan2016green}.
These techniques have enabled extensive studies of orbital magnetization in a wide range of contexts: two dimensional system{s}~\cite{xiao2007valley,liu2008orbital,yan2017orbital,marrazzo2017locality,marrazzo2016irrelevance,bianco2013orbital}, spontaneous magnetization of Fe, Co, and Ni~\cite{ceresoli2010first,lopez2012wannier}, ferromagnetism in magnetic thin film{s}~\cite{miron2011perpendicular}, topological orbital ferromagnets~\cite{hanke2016role}, and perovskite transition-metal oxides~\cite{Nikolaev2014orbital}. 
{This framework has also been applied to formulate nuclear magnetic resonance~\cite{thonhauser2009converse} and electron paramagnetic resonance~\cite{rohrmuller2017electron,ceresoli2010first} as derivatives of the orbital magnetization.}

More recently, a theory of multipole moments in quantum solids has been developed to study higher electric moments, initially in the context of topological crystalline insulator{s}~\cite{benalcazar2017A,benalcazar2017B}. 
For example, in Ref.~\onlinecite{benalcazar2017A}, a two dimensional tight-binding model, having either mirror symmetries along the $x$ and $y$ axes or $C_4$ symmetry, {was} proposed to have a well-defined~\footnote{An important aspect of higher moments, both electric and magnetic, is that all the lower moments must vanish for the leading nonzero moment to be well-defined and origin-independent~\cite{raabbook}.}, quantized electric quadrupole moment $q_{xy}$. As a consequence, the model can exhibit quantized dipole moments tangent to its edges and fractional corner charges where two edges intersect. Soon after the initial theoretical proposal, the model was realized in a trio of metamaterial platforms
\cite{serra2018observation,peterson2018quantized,imhof2018}  where analogs of the proposed phenomena were experimentally confirmed. Despite such exciting advances in the field, a rigorous microscopic understanding of generic, \emph{non-quantized} electric multipole moments is challenging since, again, the position operator is ill-defined in a periodic  solid. However, the first steps toward a generic and many-body framework were proposed and discussed in Refs.~\onlinecite{wheeler2019many,kang2019many,ono2019difficulties,dubinkin2021theory}. Additionally, Refs.~\onlinecite{Ren2021} and \onlinecite{Trifunovic2020} showed that the edge polarizations used to determine the electric quadrupole moment are only well-defined in the presence of quantizing symmetries, and they emphasized the physical relevance of the corner charge.

As a complement to the work on electric quadrupole moments, in this work, we describe an approach to study the diagonal components of the \emph{magnetic} quadrupole moment (MQM) tensor in three-dimensional insulators. We combine calculations of hinge currents with a new slab-geometry method to calculate surface magnetization to determine the MQM (see Ref.~\onlinecite{vanderbilt2022} for a recent work proposing a similar surface magnetization formalism).  One of the key observable signatures of these MQMs is the presence of hinge currents localized at the intersection of two surfaces.
Similar to the corner charge found in the electric quadrupole insulator, the hinge current derived from a bulk MQM is a bound current that is \emph{shared} by the magnetic dipole moments localized on the adjacent surfaces. We demonstrate that, analogously to the electric case, a bulk MQM can be defined as a linear combination of of boundary observables, namely surface magnetizations and hinge currents. 

Naively, from the perspective of classical electromagnetism, one would expect the hinge currents to be purely determined by the difference of the surface magnetizations of the surface that meet at the hinge. However, here we explicitly demonstrate that the bulk MQM quantifies the amount by which the \emph{full} hinge current deviates from this expectation. This is analogous to the electric quadrupole case in two dimensions, where the mismatch between the corner charge and the sum of the polarizations of the edges meeting at the corner is precisely quantified by the bulk electric quadrupole moment. Furthermore, while one can determine the bulk MQM of a system from its boundary signatures, we show that it cannot be changed by surface modifications, e.g.,  attaching lower-dimensional phases to system boundaries, as long as the total magnetic moment remains zero. In this sense, the MQM captures a bulk property of a 3D system. 

Remarkably, we show that the change in the MQM as a function of chemical potential can be a topological invariant that identifies higher order topology under certain symmetry conditions. To this end, for our example systems we concentrate on the MQM in three-dimensional higher order topological insulators (HOTIs)~\cite{benalcazar2017B,schindler2018higher}. Indeed, despite having a gapped bulk and gapped surfaces, such phases can host gapless modes localized on the hinges between adjoining surfaces. In our case we consider time-reversal breaking HOTIs that exhibit chiral hinge modes. We will see that these modes can be a signature of a nonzero bulk MQM in cases where the total magnetic moment of the HOTI vanishes.
Explicitly, we test our formalism in two examples of HOTIs, one intrinsic and one extrinsic. Intrinsic HOTIs have anomalous boundaries that cannot be trivialized by attaching lower-dimensional phases, while boundary signatures of extrinsic HOTIs~\cite{benalcazar2017A,Geier2018} can be removed by adding surface layers. The MQM itself is invariant under such attachment, and thus can describe features of both types of HOTIs and other phases including boundary obstructed topological phases (BOTPs)\cite{khalaf2019boundary}. 

As mentioned above, for the majority of this article we restrict our attention to the diagonal components of the bulk MQM and argue that they can be determined as linear combinations of the surface-localized magnetizations and hinge currents. 
Previous studies that include or focus on the off-diagonal MQM components have appeared in Refs.~\onlinecite{shitade2018theory,gao2018orbital,winkler2020collinear}. We discuss connections to these works in Sec. \ref{sec:Discussion}, but we will leave an adaptation of our methods to the off-diagonal components to future work. 

This paper is organized as follows. 
In Sec.~\ref{sec: Magnetic quadrupole moments in a finite system}, we review some features of the magnetic dipole moment and MQM in the classical continuum. Subsequently, we formulate a definition of the MQM in terms of its boundary manifestations. 
In Sec.~\ref{sec:Surface magnetization on the slab geometry}, 
we propose a method to calculate  layer-resolved magnetization in a slab geometry, from which a general quantum mechanical formula is obtained for the diagonal components of the (orbital) MQM. 
In Sec.~\ref{sec:MQM in higher order topological phases} we consider examples of HOTIs with nonzero bulk MQM and find derivatives of its components can serve as topological invariants in such systems. In Sec.~\ref{sec:symmetry} we demonstrate how spatial symmetries generically constrain the components of the MQM and their associated derivative topological invariants.
In Sec.~\ref{sec:Discussion}, we discuss the connection between this work and recent work mentioned above~\cite{shitade2018theory,gao2018orbital} that also presents a framework for calculating the MQM. We note that these works do not yield non-vanishing values of the MQM for the HOTIs we study here, and we address this discrepancy.  We also briefly discuss the off-diagonal components of the MQM. Finally, we conclude in Sec.~\ref{sec:Conclusion}. 
In App.~\ref{sec:Multipole expansion for vector potential and boundary manifestation of magnetic multipoles}, we review the multipole expansion of the vector potential, from which the magnetic multipoles and their boundary manifestations are defined. 
In App.~\ref{sec:Condition_zero_M}, we discuss the conditions for a well-defined MQM.

\section{Magnetic quadrupole moments in a finite system}
\label{sec: Magnetic quadrupole moments in a finite system}

Generally, it is a nontrivial task to directly calculate magnetic multipole moments in a periodic lattice because the position operator is ill-defined. 
Consequently, it is difficult to connect the macroscopic observables to the microscopic theory for the MQM. Instead of presenting a microscopic theory, in this section we study the macroscopic signatures of the MQM, namely hinge currents and surface magnetization, for a classical continuum. 
Similar to the electric quadrupole moment~\cite{benalcazar2017A, benalcazar2017B}, we will argue that the MQM, as a bulk property, can be determined from a combination of boundary responses, i.e., the hinge currents and surface magnetizations.

\subsection{Review of magnetic dipole moment}
\label{sec:Review of magnetic dipole moment}
In this subsection we briefly review the classical continuum formulation of the magnetization, as well as the theory of orbital magnetization for quantum tight-binding models.
It is instructive to first consider the magnetic dipole moment in a classical {continuous} medium~\cite{raabbook}:
\begin{eqnarray}\begin{aligned}
\label{eq:def_MDM}
{\bf M} = \frac{1}{2V}\int {\bf r}\times {\mathbfcal J} dV,
\end{aligned}\end{eqnarray}
where ${\bf r}$ is the position coordinate, ${\mathbfcal J}=\mathcal{J}_1\hat{x}+\mathcal{J}_2\hat{y}+\mathcal{J}_3\hat{z}$ is the current density, and $V$ is the sample volume. We henceforth take the signed electric charge $e<0$ and set its magnitude to unity in what follows.
In a fully open geometry, there is a bound, macroscopic current
\begin{eqnarray}\begin{aligned}
\label{eq:edge_current_MDM}
{\bf J}^\alpha = {\bf M}\times{\bf n}^\alpha
\end{aligned}\end{eqnarray}
circulating around the sample on the surface normal to ${\bf n}^\alpha$. Here the Greek letters  $\alpha,\beta=\pm x,\pm y, \pm z$ label the surfaces with outward normal vectors ${\bf n}^\alpha$ having components
{
\begin{eqnarray}\begin{aligned}
\label{eq:def_n}
n_i^\alpha=s_\alpha\delta^{|\alpha|}_i,
\end{aligned}\end{eqnarray}
}where the sign $s_{\alpha=\pm}=\pm1$ encodes the orientation~\cite{benalcazar2017A}.

For a tight-binding Hamiltonian on a lattice, we treat ${\bf r}$ as an operator, and replace the current density $\mathbfcal{J}$ with the velocity operator 
\begin{eqnarray}\begin{aligned}
\label{eq:def_v}
\hat{\bf v} = i\left[\hat{H},\hat{\bf r}\right], 
\end{aligned}\end{eqnarray}
{where $\hat{H}$ is the Hamiltonian of the system. The integral in Eq.~\eqref{eq:def_MDM} is replaced by a sum over the occupied eigenstates such that}
\begin{eqnarray}\begin{aligned}
\label{eq:MDM_lattice}
{\bf M} = -\frac{1}{2V}\sum_{n\in\text{occ}}\langle w_n|(\hat{\bf r}\times \hat{\bf v})|w_n\rangle,
\end{aligned}\end{eqnarray}
where $|w_n\rangle$ are the occupied eigenstates of the given Hamiltonian $\hat{H}$.
One can employ Eq.~\eqref{eq:MDM_lattice} to calculate magnetic dipole moment in finite open systems, where $\hat{\bf r}$ is diagonal in a tight-binding basis.
However, the position operator becomes {ill-}defined when the system has an infinite spatial extent or periodic boundary conditions. This problem also occurs for electric polarization~\cite{resta1998quantum}, and for the magnetization it {has} been solved to arrive at a ``modern theory of magnetization'' ~\cite{thonhauser2005orbital,xiao2005berry,xiao2006berry,Ceresoli2006,shi2007quantum}. The results of this work determine the orbital magnetization to be
\begin{eqnarray}\begin{aligned}
\label{eq:MDM}
{\bf M} = \frac{1}{2}\text{Im} \hspace{-.2cm} \sum_{n\in\text{occ}, {\bf k}}\langle\partial_{\bf k}u_{n \bf k}|\times (H_{\bf k}+\epsilon_{n\bf k}-2\mu)|\partial_{\bf k}u_{n \bf k}\rangle,
\end{aligned}\end{eqnarray}
where $H_{\bf k}\equiv e^{-i{\bf k}\cdot{\bf r}}\hat{H}e^{i{\bf k}\cdot{\bf r}}$ is the Bloch Hamiltonian, $\mu$ is the chemical potential, {$|u_{n\bf k}\rangle$} is the periodic part of the Bloch wave function of the $n$-th band, which satisfies {$H_{\bf k}|u_{n \bf k}\rangle=\epsilon_{n\bf k}|u_{n \bf k}\rangle$}, and the summation runs over all the occupied bands and occupied momenta in the Brillouin zone. We note that Eq.~\eqref{eq:MDM} has also been generalized to finite temperature and interacting system{s} in Ref.~\onlinecite{shi2007quantum}. 

Since it will be useful later, let us mention that 2D systems with nonzero magnetic dipole moment must break time reversal (TR) symmetry and mirror symmetries, as is evident from Eq.~\eqref{eq:def_MDM}. Two canonical examples are 2D Dirac Semimetals (DSMs) with nodes offset in energy~\cite{ramamurthy2015patterns}, and Chern insulators. In both cases the magnetic dipole moment manifests as circulating edge current, as expected from Eq.~\eqref{eq:edge_current_MDM}. We will elaborate on this point in Sec.~\ref{sec:MQM in higher order topological phases}, as our intuition for magnetization in 2D helpful for our understanding of the MQM in higher order topological phases, and in particular its boundary manifestations. To this end, let us {now move on to derive} the relationship between the hinge current and {the} bulk MQM.

\subsection{Classical definition of magnetic quadrupole moment}
\label{sec:Classical definition of magnetic quadruple moments}

\begin{figure}
\centering
\includegraphics[width=0.8\linewidth]{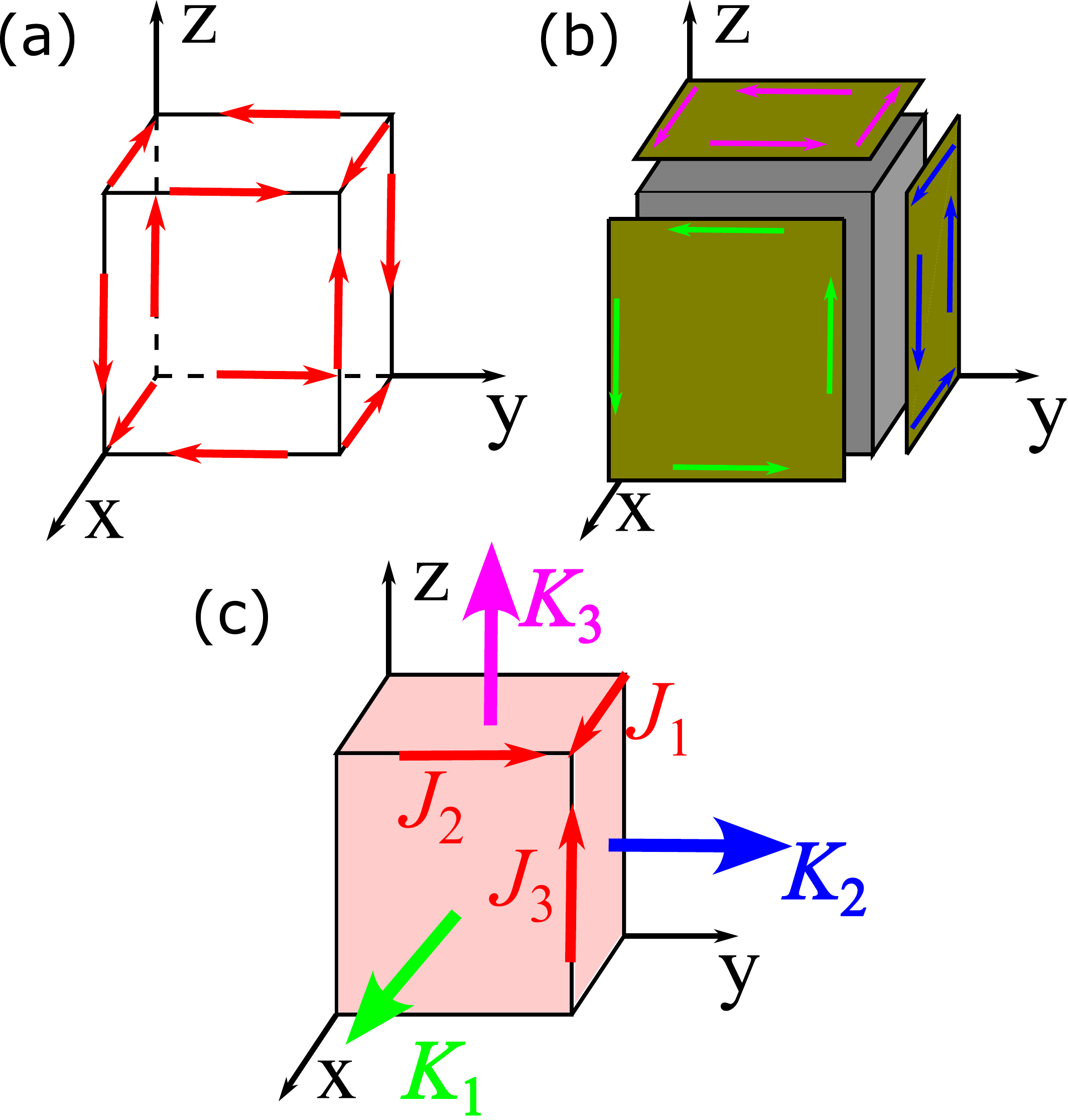}
\caption{(a) A  configuration of hinge currents {with vanishing total magnetization}. The arrows show our convention for the positive directions of the hinge currents according to Eq.~\eqref{eq:Hinge_currents}, which are opposite for opposite hinges related by $C_{2i}$ symmetry. At the corner, the current follows the conservation relation in Eq.~\eqref{eq:Conservation_currents}. (b) Decorating a trivial 3D insulator (grey) with 2D Chern insulators can also produce  hinge currents but no MQM. (c) Boundary observables for an orthorhombic crystal used to define the MQM in Eq.~\eqref{eq:MQM_as_J_K}.}
\label{fig:1}
\end{figure}

Magnetic multipole moments are defined via the multipole expansion of the vector potential~\cite{raabbook}. As we explicitly derive in App.~\ref{sec:Multipole expansion for vector potential and boundary manifestation of magnetic multipoles}, the MQM can be defined in the continuum as
\begin{eqnarray}\begin{aligned}
\label{eq:classical_MQM}
M_{ij} = \frac{2}{3V}\int_V r_i({\bf r}\times \mathbfcal{J} )_jdV.
\end{aligned}\end{eqnarray}
Much like the electric {quadrupole moment}, {the} MQM is well-defined only if the bulk magnetic dipole moment is zero. To see this, consider shifting the coordinate origin ${\bf r}\rightarrow {\bf r}+{\bf a}$ by a constant vector ${\bf a}$. As a result, the MQM transforms as
\begin{align}
\label{eq:shift_MQM}
M_{ij}&\rightarrow M_{ij} -\frac{2}{3}a_iM_j-\frac{2}{3}\delta_{ij}{\bf a}\cdot {\bf M} \nonumber\\ &+ \frac{2}{3V} a_i({\bf a}\times\int_V\mathbfcal{J}dV )_j,
\end{align}
where we used $V^{-1}\int r_i\mathcal{J}_jdV=\epsilon_{ijk}M_k$ (c.f. App.~\ref{sec:Some useful identities}). Since $\int \mathbfcal{J}dV=0$ {at equilibrium}, and ${\bf a}$ is an arbitrary displacement, we find that the MQM, as a rank-2 tensor, is origin-independent if and only if ${\bf M}=0$. This is a general and well-known result: only the leading non-vanishing electric or magnetic moments are origin independent and well-defined. 
Despite the similarity, we note two important features of {the} MQM that are very different from the electric quadrupole moment. First, {$M_{ij}$} is \emph{not} {a} symmetric tensor, and the off-diagonal components have different physical meanings from each other, as will be discussed further in Sec.~\ref{sec:Discussion}.
Second, {the} MQM is a traceless tensor, as evident from Eq.~\eqref{eq:classical_MQM}. As we will see, this provides strong constraints on the possible form of {the} diagonal components of the MQM, which are the primary focus of this paper. Finally, we note that, some components of the MQM can still be well-defined even if the magnetization is non-vanishing in one direction. See Sec.~\ref{sec:Discussion} for a brief additional discussion of this point. 

In order to have a well-defined MQM we want to generically fix the magnetization to vanish. For crystalline systems, the bulk orbital magnetization can be constrained to be zero by symmetries. From Eq.~\eqref{eq:def_MDM}, we notice that under a two-fold rotation along the $\hat{z}$-direction, the $z$-component of the magnetization remains invariant while the other two change sign. Applying the same reasoning to the other directions, it follows that the magnetization of a crystalline system vanishes if it has the two-fold rotation symmetries $C_{2x}$, $C_{2y}$, and $C_{2z}$. Hence, we will focus on orthorhombic crystals since they generically have these symmetries.

With this setup we are now ready to determine the boundary phenomena associated with the MQM. The diagonal components of the bulk MQM manifest on the boundary as surface localized magnetizations and localized currents on the hinges where surfaces intersect. As we derive in App.~\ref{sec:Multipole expansion for vector potential and boundary manifestation of magnetic multipoles}, the hinge currents and surface magnetizations resulting from a non-vanishing MQM are
\begin{align}
    J^{\alpha\beta}_i &=\frac{1}{2}\epsilon_{ikl}(n^{\alpha}_jn^{\beta}_lM_{jk}+n^{\beta}_jn^{\alpha}_lM_{jk})\label{eq:Hinge_currents},\\
    K^\alpha_j &= n^\alpha_kM_{jk}\label{eq:surface_M},
\end{align}
where $\alpha,\beta=\pm x,\pm y,\pm z$ label the surfaces{ with outward normal vectors $n_i^\alpha$ as defined in Eq.~\eqref{eq:def_n}}. 
Here $J_l^{\alpha\beta}$ and $J_l^{\beta\alpha}$ are the same quantity, namely the current on the hinge between surfaces $\alpha$ and $\beta$ (see below).
$K^\alpha_j$ is the $j$-th component of the magnetization ${\bf K}^\alpha$ localized on the surface $\alpha$.

Given the important role of hinge currents in our later discussion, {let us} consider an explicit geometry to properly understand Eq.~\eqref{eq:Hinge_currents}. In Fig.~\ref{fig:1}(a),
we show a configuration of hinge currents for an orthorhombic crystal. The red arrows are not the pattern of hinge currents, but instead correspond to \emph{our convention} for the positive direction of the currents on each hinge. To simplify later definitions, we have chosen this convention so that positive currents are mapped to positive currents by the $C_{2x}, C_{2y},$ and $C_{2z}$ symmetries of the orthorhombic lattice.
By current conservation, we then have
\begin{eqnarray}\begin{aligned}
\label{eq:Conservation_currents}
J^{+y,+z}_1+J^{+z,+x}_2+J^{+x,+y}_3=0,
\end{aligned}\end{eqnarray}
at the corner shared by $+\hat{x}$, $+\hat{y}$ and $+\hat{z}$ surfaces. Similar relation{s} hold for other corners.
As shown in App.~\ref{sec:Condition_zero_M}, not every configuration of hinge currents allows for a well-defined MQM as some configurations have non-vanishing magnetization. {To ensure that the system has zero magnetization the currents on parallel hinges must have the same magnitudes, \emph{e.g.} ${J_3^{+x,+y}=J_3^{-x,+y}=J_3^{+x,-y}=J_3^{-x,-y}}$. Under the assumption that the hinge currents become completely localized on the hinges in the thermodynamic limit, in  App.~\ref{sec:Condition_zero_M}  we prove that this relation follows from requiring vanishing magnetization and the corner current conservation from Eq.~\eqref{eq:Conservation_currents}.} We note that this relationship is automatically satisfied as a consequence of the $C_{2i}$ symmetries for orthorhombic crystals.  As such, given such $C_{2i}$ symmetries, the boundary observables, including hinge currents and surface magnetizations, are no longer independent. In fact, all boundary observables can be determined from three orthogonal $J_i^{\alpha\beta},$ and three orthogonal $K_j^\alpha$. For simplicity we will often leave off the superscripts and consider the six quantities $J_{1,2,3}$ and $K_{1,2,3}$  as defined (with arrows indicating their positive orientations) in Fig.~\ref{fig:1}(c). 

\subsection{Magnetic quadrupole moment as a bulk property}
\label{sec: Magnetic quadrupole moment as a bulk property}

Although hinge current is one of the boundary manifestations of the bulk MQM, it can also arise purely from surface decoration. Thus, to illustrate the bulk nature of the quadrupole moment, we must distinguish between systems with hinge currents induced by a bulk MQM, and those with hinge currents induced by a surface magnetization in the absence of MQM.
To see that hinge currents can arise without a bulk MQM, we consider a trivial system with $M_{ij}=0$ and glue 2D magnetic materials, such as Chern insulators, on its surfaces, as shown in Fig.~\ref{fig:1}(b). 
This induces a surface localized magnetization ${ \tilde{\bf K}}^\alpha$ on the surface normal to ${\bf n}^\alpha$. Following Eq.~\eqref{eq:edge_current_MDM}, the current on the hinge connecting two adjacent surfaces $\alpha$ and $\beta$ reads
\begin{eqnarray}\begin{aligned}
\label{eq:hinge_current_decoration}
{\tilde{\bf J}}^{\alpha\beta}={ \tilde{\bf K}}^\alpha\times{\bf n}^\beta + {\bf \tilde{ K}}^\beta\times{\bf n}^\alpha.
\end{aligned}\end{eqnarray}
In terms of components, we have
\begin{align}
\label{eq:amperes_law}
    \tilde{J}^{+x,+y}_{3} &= \tilde{K}^{+x}_1 - \tilde{K}^{+y}_2, \nonumber\\
    \tilde{J}^{+y,+z}_{1} &= \tilde{K}^{+y}_2 - \tilde{K}^{+z}_3, \\
    \tilde{J}^{+z,+x}_{2} &= \tilde{K}^{+z}_3 - \tilde{K}^{+x}_1, \nonumber
\end{align}
where the hinge current is simply the difference of the magnetizations on the adjacent surfaces. This is a natural consequence of Ampère's law from classical electromagnetism, given that the bulk of the material is trivial and has no magnetism~\cite{zhu2020quantized}. 

In contrast, let us now consider adding surface layers to a system that has a non-vanishing bulk MQM. Because of the decorations, Eqs.~\eqref{eq:Hinge_currents}-\eqref{eq:surface_M} will be modified as
\begin{align}
    J^{\alpha\beta}_i &=\frac{1}{2}\epsilon_{ikl}(n^{\alpha}_jn^{\beta}_lM_{jk}+n^{\beta}_jn^{\alpha}_lM_{jk}) \nonumber\\
    &\quad+ \epsilon_{ikl}(\tilde{K}^\alpha_{k}n_l^\beta +\tilde{K}^\beta_{k}n_l^\alpha)\label{eq:Hinge_currents_modified},\\
    K^\alpha_j &= n^\alpha_kM_{jk}+\tilde{K}_{j}^\alpha\nonumber,
\end{align}
where we see that hinge currents and surface magnetizations receive bulk and surface contributions.  In order to extract the bulk MQM, we combine Eq.~\eqref{eq:Hinge_currents_modified} to eliminate $\tilde{\bf K}$, the magnetizations from the surface decoration, and arrive at
\begin{align}
\label{eq:MQM_as_J_K}
    \mathcal{M}_{12}\equiv \frac{1}{2}(M_{11}-M_{22}) &= K_1^{+x}-K_2^{+y}-J_3^{+x,+y},\nonumber\\
    \mathcal{M}_{23}\equiv \frac{1}{2}(M_{22}-M_{33}) &= K_2^{+y}-K_3^{+z}-J_1^{+y,+z},\\
    \mathcal{M}_{31}\equiv \frac{1}{2}(M_{33}-M_{11}) &= K_3^{+z}-K_1^{+x}-J_2^{+z,+x}.\nonumber
\end{align}

This set of equations is the main result of this section: it defines the \emph{bulk} MQM as a linear combination of the total hinge currents and surface magnetizations \emph{regardless} of their physical origin. 
Importantly, we cannot isolate single diagonal components like $M_{11}$ or $M_{22}$ from the effects of surface decoration. Hence,  Eq.~\eqref{eq:MQM_as_J_K} suggests that the classical MQM tensor $M_{ij}$ is not itself a well-defined bulk quantity, but instead one should consider their \emph{differences}, namely the quantities $\mathcal{M}_{ij}$ introduced above as bulk quantities. 
Indeed, the $M_{ij}$ tensor of a material is sensitive to the surface decoration, as is evident from Eq.~\eqref{eq:Hinge_currents_modified}, but $\mathcal{M}_{ij}$ remains invariant even if one arbitrarily changes the surface magnetization of the material via surface decoration. 
Furthermore, as we explain in Sec.~\ref{sec:Bulk formula for MQM}, the combinations in $\mathcal{M}_{ij}$ also naturally appear when considering linear response theory for the MQM. 
The quantity $\mathcal{M}_{ij}$ therefore characterizes a well-defined bulk magnetic property, and we hereafter refer it as the MQM. 

The definition of MQM in Eq.~\eqref{eq:MQM_as_J_K} is analogous to the definition of electric quadrupole moment in two dimensions. Recall that in Refs.~\onlinecite{benalcazar2017A, benalcazar2017B}, the electric quadrupole moment is defined as
\begin{align}
    q_{12}\equiv P_1 + P_2 - Q_\text{cor},
\end{align}
where $P_i$ are the edge polarizations for the edges normal to $\hat{i}$, and $Q_\text{cor}$ is the corner charge {shared} by the two edges. For a system with $q_{12}=0$, the corner charge is simply the sum of the polarizations, 
whereas a nonzero bulk electric quadrupole creates a mismatch between the corner charge and the edge polarization sum. 
Similarly, we have just demonstrated that a bulk magnetic quadrupole creates a mismatch between the hinge currents and the surface magnetizations. In contrast, for a system with $\mathcal{M}_{ij}=0$, the magnitude of, say, $J_3^{+x, +y}$ can be inferred by calculating the surface magnetizations on the surfaces normal to $+\hat{x}$ and $+\hat{y}$ and taking their difference.
However, this is not the case if $\mathcal{M}_{ij}\neq0$, as the bulk MQM also contributes to the hinge currents, much like the electric quadrupole moment contributions to the corner charge.
In other words, the bulk MQM quantifies the mismatch between the hinge current and the difference of the neighboring surface magnetizations, as shown in Eq.~\eqref{eq:MQM_as_J_K}.

In materials having crystal symmetry the electric quadrupole moment can be quantized and can act as a topological invariant for crystalline insulators\cite{benalcazar2017A}. It is thus natural to ask if the MQM can also serve as a topological invariant. Conventional manifestations of quadrupole moment are edge polarization and corner charge, while surface magnetic moment and hinge currents are natural manifestations of $\mathcal{M}_{ij}.$ However, magnetic moments and currents are not entirely analogous to polarizations and charges because the former lack an inherent discreteness, and are not quantized by crystal symmetries. Moreover, if one can continuously fine-tune a model Hamiltonian to have particle-hole symmetry both surface magnetizations and hinge currents are zero at half-filling, causing the bulk MQM to vanish. Both of these problems are resolved by considering not the bulk MQM itself, but rather its derivative with respect to chemical potential. Indeed, in Sec. \ref{sec:MQM in higher order topological phases} we find cases where $d\mathcal{M}_{ij}/d\mu$ are quantized. This is analogous to how Chern insulators have a continuously varying 2D magnetization arising from their chiral edge modes, yet the derivative of the magnetization with respect to the chemical potential gives a quantized invariant proportional to the Chern number\cite{Ceresoli2006}. 

In Sec.~\ref{sec:MQM in higher order topological phases}, we highlight two examples of HOTIs that can support a nonzero MQM. Our formulation of MQM provides a method to calculate it through the macroscopic quantities described above. We then show that the derivative of the MQM with respect to the chemical potential can serve to distinguish different topological phases for these HOTIs. As we will see in Sec. \ref{sec:symmetry}, much like the electric case, crystalline symmetries play a critical role in the constraints on these invariants. 

\section{Surface magnetization in a slab geometry}
\label{sec:Surface magnetization on the slab geometry}

As discussed in Sec.~\ref{sec: Magnetic quadrupole moment as a bulk property}, surface magnetization plays an essential role in our definition of MQM. Here we derive a formula for the layer-resolved magnetization $M_i(r_i)$ of a system in a slab geometry that has open boundaries in one direction and periodic boundaries in the other two directions. The surface magnetization can be extracted by summing over the contribution to the magnetization that is localized to the surface of interest.

We start by re-writing the 2D magnetization in Eq.~\eqref{eq:MDM} in position space as~\cite{bianco2013orbital}
\begin{eqnarray}\begin{aligned}
\label{eq:MQM_real_space_1}
M_i = \sum_{\bf r}\frac{1}{2}\epsilon_{ijl}\text{Im}\langle{\bf r}|& \left\{\hat{P}\hat{r}_j\hat{Q}(\hat{H}-\mu\hat{N})\hat{Q}\hat{r}_l\hat{P}\right.\\
&\left.+\hat{Q}\hat{r}_l\hat{P}(\hat{H}-\mu\hat{N})\hat{P}\hat{r}_j\hat{Q}\right\}|{\bf r}\rangle,
\end{aligned}\end{eqnarray}
where 
\begin{eqnarray}\begin{aligned}
\label{eq:def_P_Q}
\hat{P} &=\frac{1}{\Omega}\sum_{{\bf k},n\in\text{occ}}|\psi_{n{\bf k}}\rangle\langle\psi_{n{\bf k}}|, \quad\hat{Q} =\mathbb{I}-\hat{P},
\end{aligned}\end{eqnarray}
are projection operators onto the occupied and unoccupied subspaces, respectively.
Here $|{\bf r}\rangle$ is the position basis, and $\Omega$ is the volume of the 2D BZ. 
%
We also note that the operator $\hat{P}\hat{r}_j\hat{Q}$ is well-defined in periodic systems because it is invariant under shifts of the origin of coordinates. 

The summand in Eq.~\eqref{eq:MQM_real_space_1} is manifestly the \emph{local} expression for the magnetization in real space. 
More compactly, we may write 
\begin{eqnarray}\begin{aligned}
\label{eq:MQM_real_space_2}
{\bf M}({\bf r}) = \frac{1}{2}\text{Im tr} \left\{\tilde{P}_{\bf r}[\hat{P}\hat{\bf r}\hat{Q}\times \hat{Q}\left\{\hat{F},\hat{\bf r}\right\}\hat{P}]\right\},
\end{aligned}\end{eqnarray}
where 
$\left\{\cdot,\cdot\right\}$ is the anti-commutator. 
Here $\hat{F}\equiv\hat{H}-\mu\hat{N}$ is the free energy operator and 
$\tilde{P}_{\bf r}\equiv|{\bf r}\rangle\langle{\bf r}|$ is the projection operator onto position ${\bf r}$. 
The passage from Eq.~\eqref{eq:MQM_real_space_1} to  Eq.~\eqref{eq:MQM_real_space_2} can makes use of the cyclic property of the trace and the identities $\left[\hat{Q},\hat{F}\right]=\left[\hat{P},\hat{F}\right]=0$.

For clarity, let us specialize to the evaluation of $M_3(z)$ in an $N_z$-layer slab geometry, where $k_{x,y}$ remain good quantum numbers because of the periodic boundary conditions in $x$ and $y$. In this case, the system can be imagined as a 2D system consisting of ``super-cells'' that are $L_z$ times larger than the original unit cells, and the definitions of the projection operators $\hat{P}$ and $\hat{Q}$ in Eq.~\eqref{eq:def_P_Q} remain valid, with ${\bf k}=(k_x,k_y)$. Though we have open boundary conditions in the $\hat{z}$-direction, we can nevertheless expand the first term in  Eq.~\eqref{eq:MQM_real_space_1} as 
\begin{eqnarray*}\begin{aligned}
&\text{Im tr}\left\{\tilde{P}_{\bf r}\hat{P}\hat{x}\hat{Q}(\hat{H}-\mu\hat{N})\hat{Q}\hat{y}\hat{P}\right\}\\
=&\frac{1}{\Omega^4}\text{Im tr}\sum_{{\bf k_1,k_1'},n,n'\in\text{occ}}\sum_{{\bf k_2,k_2'},m,m'\in\text{unocc}}\tilde{P}_{\bf r}|\psi_{n{\bf k}_1}\rangle\langle\psi_{n{\bf k}_1}|\\
&\hat{x}|\psi_{m{\bf k}_2}\rangle\langle\psi_{m{\bf k}_2}|(\hat{H}-\mu\hat{N})|\psi_{m'{\bf k}_2'}\rangle\langle\psi_{m'{\bf k}_2'}|\hat{y}|\psi_{n'{\bf k}_1'}\rangle\langle\psi_{n'{\bf k}_1'}|\\
=&-\text{Im tr}\sum_{{\bf k}_1,nn'\in\text{occ},m\in\text{unocc}}\tilde{P}_{\bf r}|u_{n{\bf k}}\rangle\langle u_{n{\bf k}}|\partial_xu_{m{\bf k}}\rangle\times\\
&\quad\langle u_{m{\bf k}}|(H_{\bf k}-\mu)|u_{m{\bf k}}\rangle\langle u_{m{\bf k}}|\partial_yu_{n'{\bf k}}\rangle\langle u_{n'{\bf k}}|\\
=&\text{Im}\sum_{{\bf k}}\tilde{P}_{\bf r}\partial_xP_{\bf k}Q_{\bf k}(H_{\bf k}-\mu)Q_{\bf k}\partial_yP_{\bf k},
\end{aligned}\end{eqnarray*}
where we used 
$\langle\psi_{n{\bf k}}|\hat{r}_j|\psi_{m{\bf k}'}\rangle=i\Omega\langle u_{n{\bf k}}|\partial_ju_{m{\bf k}'}\rangle\delta_{\bf kk'}$ for $n\neq m$~\cite{rauch2018geometric}. Here $P_{\bf k}\equiv \sum_{n\in\text{occ}}|u_{n\bf k}\rangle\langle u_{n\bf k}|$ and $Q_{\bf k}\equiv \sum_{m\in\text{unocc}}|u_{m\bf k}\rangle\langle u_{m\bf k}|$ are the projection operators in terms of the periodic part of the Bloch wavefunction.
Similar gymnastics can be carried out for the second term as well, and we arrive at
\begin{eqnarray}\begin{aligned}
&M_3({\bf r}) = \frac{1}{2}\text{Im tr}\sum_{\bf k}\left\{\tilde{P}_{\bf r}(g_{{\bf k},xy}+h_{{\bf k},xy})-(x\leftrightarrow y)\right\},
\end{aligned}\end{eqnarray}
where 
\begin{eqnarray}\begin{aligned}
\label{eq:def_g_h_nk}
g_{{\bf k},jl} &=\partial_jQ_{\bf k}Q_{\bf k}(H_{\bf k}-\mu)Q_{\bf k}\partial_lQ_{\bf k},\\
h_{{\bf k},jl} &=\partial_lP_{\bf k}P_{\bf k}(H_{\bf k}-\mu)P_{\bf k}\partial_jP_{\bf k}.
\end{aligned}\end{eqnarray}
Upon taking the limit $L_{x,y}\rightarrow\infty$, the layer resolved $M_3(z)$ becomes
\begin{eqnarray}\begin{aligned}
\label{eq:Mz_zero_T}
M_3(z) &\equiv \frac{1}{L_xL_y}\int dxdyM_z({\bf r}) \\
&=\frac{1}{2}\sum_{\bf k}\text{Im tr}\left\{\tilde{P}_{z}(g_{{\bf k},xy}+h_{{\bf k},xy})-(x\leftrightarrow y)\right\},
\end{aligned}\end{eqnarray}
where 
\begin{eqnarray}\begin{aligned}
\tilde{P}_z &=\frac{1}{L_xL_y}\int dxdy \tilde{P}_{\bf r}
\end{aligned}\end{eqnarray}
is the operator that projects onto the $z$-th layer. 

This can be generalized to other components as
\begin{eqnarray}\begin{aligned}
\label{eq:layer_resolved_MDM}
M_i(r_i) &=\frac{1}{2}\epsilon_{ijl}\sum_{\bf k_\perp}\text{Im tr}\left\{\tilde{P}_i(g_{{\bf k_\perp},jl}+h_{{\bf k_\perp},jl})\right\},
\end{aligned}\end{eqnarray}
where the index $i$ is not summed over, and ${\bf k_\perp}$ is used to denote the momenta transverse to $r_i$. 
Given this explicit expression for the layer resolved magnetization, we can finally define the surface magnetization as
\begin{eqnarray}\begin{aligned}
\label{eq:K_as_M}
K_i\equiv \int_0^{L_i/2}dr_iM_i(r_i),
\end{aligned}\end{eqnarray}
for a slab of thickness $L_i$ centered at $0$, with open boundary conditions along the $\hat{i}$-direction, and periodic boundary conditions along the other two transverse directions. 
We will now explicitly make use of Eqs.~\eqref{eq:layer_resolved_MDM}, \eqref{eq:K_as_M}, together with Eq.~\eqref{eq:MQM_as_J_K},  to determine the MQM for various higher order topological phases in Sec.~\ref{sec:MQM in higher order topological phases}. Furthermore, we will show that derivatives of the MQM with respect to the chemical potential can act as topological invariants in some higher order phases.  

\section{MQM and Topological Invariants in higher order topological insulators}
\label{sec:MQM in higher order topological phases}

We have shown that the MQM, as defined in Eq.~\eqref{eq:MQM_as_J_K}, is a bulk quantity that can be calculated by combining surface magnetizations and hinge currents. 
We also recall that the magnetization of a system must be zero in order to support a well-defined MQM.
This can be achieved by imposing the twofold rotation symmetries $C_{2x}$, $C_{2y}$, and $C_{2z}$, as is clear from Eq.~\eqref{eq:def_MDM}. 
In addition, time-reversal symmetry must be broken for the MQM to be nonzero. 
In this section, we present two examples of HOTIs, one intrinsic and one extrinsic, which have these symmetry properties and support chiral hinge states and nontrivial MQMs.

\subsection{MQM in an intrinsic HOTI}
\label{sec:MQM in HOTIs}

We first consider an intrinsic HOTI with the tight-binding Hamiltonian~\cite{schindler2018higher}
\begin{equation}\label{eq:c4t_ham}
\begin{aligned}
H^\text{int}(\bf{k}) =& (M+t\sum_i \cos k_i ) \tau_3 \sigma_0 + \Delta_1 \sum_i \sin k_i \, \tau_1 \sigma_i \\ 
&+ \Delta_2 (\cos k_x  - \cos k_y ) \, \tau_2 \sigma_0,
\end{aligned}
\end{equation}
where $\tau_i$ and $\sigma_i$ act on orbital and spin degrees of freedom, respectively. 
This model has two-fold rotation symmetries with the representations 
\begin{align}
\label{eq:2_fold_symmetries}
    C_{2x} &=\tau_0\sigma_1,\nonumber\\
    C_{2y} &=\tau_0\sigma_2,\\
    C_{2z} &=\tau_0\sigma_3.\nonumber
\end{align}
Consequently, $H^\text{int}(\bf{k})$ is guaranteed to have zero magnetization and a well-defined MQM, which we will see below is non-vanishing. To proceed, we note that when $1<|M/t|<3$ and $\Delta_2=0$, the system is in fact a first-order 3D topological insulator with surface Dirac cones protected by time-reversal symmetry $T=\tau_0 \sigma_2 K$. In addition to TR symmetry, when $\Delta_2=0$ the system is also invariant under fourfold rotations around the $z$-axis with ${C_{4z}} = \tau_0 e^{-i\frac{\pi}{4} \sigma_3}$.  

\begin{figure}
    \hspace{1.5cm}
    \includegraphics[width=0.7\linewidth]{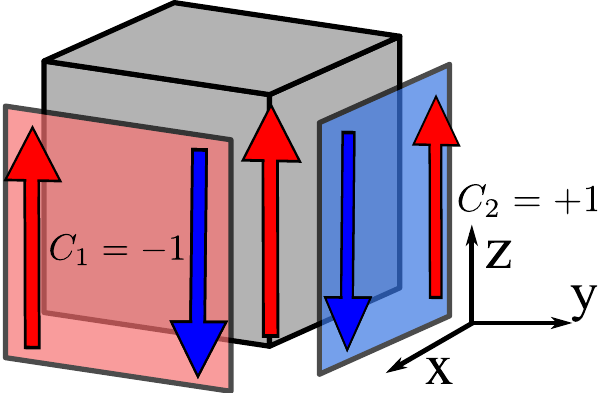}
	\caption{A $C_{4z}T$ HOTI in a nontrivial phase with a single chiral hinge mode. One can glue 2D Chern insulators to its surfaces in a $C_{4z}T$-symmetric fashion to change the direction of the hinge current. This process, however, will not change the bulk MQM of the system.}
    \label{fig:2}
\end{figure}

The model in \eqref{eq:c4t_ham} cannot support a nonzero MQM when $\Delta_2=0$ because it is $T$-symmetric.
Taking $\Delta_2\neq0$ introduces a ${C_{4z}}$ and $T$-breaking mass term that gaps out the $x$- and $y$-normal surfaces and leaves gapless chiral modes traversing the $z$-parallel hinges where these surfaces intersect.
While ${C_{4z}}$ and $T$ are individually broken, the product $C_{4z}T$ remains unbroken. The composite symmetry protects the odd number of Dirac cones on the $z$-normal surface. Hence, when $1<|M/t|<3$ and $\Delta_2\neq0,$ the resulting system is an intrinsic higher-order phase. The intrinsic nature is characterized by chiral hinge modes that cannot be completely removed by gluing lower-dimensional phases to the surfaces or hinges while respecting $C_{4z}T$~\cite{Geier2018}. Furthermore, this phase can be described by the same $\mathbb{Z}_2$ topological invariant as the 3D $T$-invariant topological insulator, namely the bulk $\theta$-angle that is quantized by $T$ in the former case~\cite{qi2008topological}, and $C_{4z}T$ in the latter~\cite{schindler2018higher}.
The $\mathbb{Z}_2$ nature of the hinge modes can be understood via gluing 2D Chern insulators on the surfaces of the system. As shown in Fig.~\ref{fig:2}, in order to respect the $C_{4z}T$ symmetry and preserve the surface gaps, the attached insulators must have opposite Chern numbers on neighboring surfaces. Hence, the number of hinge modes can change only by an even number, in agreement with the $\mathbb{Z}_2$ nature of the bulk invariant. 

The fact that the number of hinge modes, and hence the hinge current, can be changed by gluing lower dimensional phases to the boundary is precisely the motivation for introducing the notion of a bulk MQM. We now argue that the derivative of the MQM encodes another topological invariant for HOTIs, 
namely the net number of chiral hinge modes \emph{arising from the bulk}:
\begin{equation}
    \label{eq:bulk_modes}
    n_{ij} \equiv 2\pi \frac{d\mathcal{M}_{ij}}{d\mu},
\end{equation}
where the hinge in question is between the surfaces normal to directions $\hat{i}$ and $\hat{j}$. 
While the net hinge current, which includes both bulk and surface contributions, changes with surface decoration (c.f. Fig. \ref{fig:2}), the bulk contribution associated with the MQM does not, as we have described in Sec.~\ref{sec: Magnetic quadrupole moment as a bulk property}. Furthermore, we show that in the presence of $C_{4z}T$ symmetry,
\begin{align}
    \label{eq:inv_C_4zT}
    n^\text{int}_{12}\in 2\mathbb{Z},~ n^\text{int}_{23}\in \mathbb{Z},~ n^\text{int}_{31}\in \mathbb{Z}, 
\end{align}
meaning that the invariant defined above always takes integer values, and the component $n_{12}$ is always an even integer for this symmetry class.
Hereafter, relations with superscript ``int'' are specific to the intrinsic HOTI, while all other relations are generic.

To demonstrate these claims, let us start with the component
\begin{align}
\label{eq:n_12}
    n_{12} = 2\pi\frac{d(K_1-K_2)}{d\mu}-2\pi\frac{d J_3}{d\mu},
\end{align}
which we claim characterizes the difference of the number of chiral and anti-chiral $z$-direction hinge modes contributed by the bulk. To show that $n_{12}$ is quantized and can be only an even integer for the intrinsic HOTI, we now show that $2\pi\frac{d(K_1-K_2)^\text{int}}{d\mu}$ and $2\pi\frac{dJ^\text{int}_3}{d\mu}$ are independently integer-quantized and always have the same parity. 

For the hinge current, we consider its contribution in a geometry with periodic boundary conditions along the $z$-direction and open boundaries in the $x$ and $y$-directions. In this geometry, the hinges host the only gapless degrees of freedom, so as long as we remain within the bulk and surface gaps, the effect of raising the chemical potential is only to fill more states on the hinge. Any change in hinge current therefore arises entirely from the hinge states. 
Following the arguments in Refs. \onlinecite{Ceresoli2006,zhu2020quantized}, each filled hinge state contributes $v(k_z)/L$ to the current, where $v(k_z)$ is the velocity of the state located at $k_z,$ and $L$ is the length of the hinge. Upon raising the chemical potential by a small amount $d\mu$, we fill an extra ${|dk_z|}/{2\pi}$ states per unit length. Given $n$ branches that cross the Fermi level, the total change in current during this process is then 
\begin{equation}\label{eq:curr_change}
dJ_3 = -\frac{1}{L}\sum_i^n L \frac{|dk_z|}{2\pi} v(k_F).
\end{equation}
Using the relationship $dk_z = d\mu/v_F$ at the Fermi level~\cite{Ceresoli2006}, we finally arrive at 
\begin{equation}
	\label{eq:curr_deriv}
	2\pi \frac{dJ_3}{d\mu} = -\sum_i^n \text{sgn}(v_F) \in \mathbb{Z}.
\end{equation}
In short, as long as $\mu$ remains inside the bulk and surface gaps, the slope of the hinge current as a function of $\mu$ is an integer that counts the net number of chiral hinge modes. 

Now, if we specialize this general result to the intrinsic HOTI considered, $2\pi \frac{dJ_3}{d\mu}$ itself can be regarded as a topological invariant, which reads
\begin{align}
\label{eq:J_3_C4zT}
    2\pi \frac{dJ_3^\text{int}}{d\mu} \in \begin{cases}
  2\mathbb{Z}+1 & \text{topological}\\   
  2\mathbb{Z} & \text{trivial}
\end{cases}.
\end{align}
This invariant corresponds to the total number of hinge modes, and includes both the contributions from the bulk and the surfaces.

To find the contribution of the surface magnetizations to the invariant $n_{12}$, note that the derivative of the surface magnetization in Eq.~\eqref{eq:layer_resolved_MDM} with respect to chemical potential gives the layer-resolved Chern number in a slab geometry~\cite{Essin2009Magnetoelectric}. The resulting surface Hall conductance is in turn proportional to the bulk $\theta$-angle, and we find
\begin{align}
    2\pi \frac{dK_1}{d\mu} = 2\pi \frac{dK_2}{d\mu} = \frac{\theta}{2\pi} \, (\text{mod  } 1).
\end{align}
The $C_{4z}T$ symmetry enforces ${K_1^\text{int}=-K_2^\text{int}}$, and also quantizes the values of $\theta$ to be $0$ or $\pi$. Consequently, for the intrinsic HOTI this expression becomes
\begin{align}
\label{eq:K_1_K_2_C4zT}
    2\pi \frac{dK_1^\text{int}}{d\mu} = -2\pi \frac{dK_2^\text{int}}{d\mu} \in \begin{cases}
  \mathbb{Z}+\frac{1}{2} & \text{topological}\\   
  \mathbb{Z} & \text{trivial}
\end{cases}.
\end{align}
Hence, $\frac{d(K_1^\text{int}-K_2^\text{int})}{d\mu}$ is an odd integer in the topological phase and an even integer otherwise.  

Now we arrive at our result. Since both hinge and surface contributions to the bulk invariant $n_{12}$ are always integers of the same parity, we finally find $n^\text{int}_{12}\in 2\mathbb{Z}$. Furthermore, we will see in Sec. \ref{sec:symmetry} that this argument applies to any system with $C_{4z}T$ symmetry and bulk and surface gaps.
Unlike the hinge current $J_3,$ or the surface magnetizations $K_1$ and $K_2$, the $2\mathbb{Z}$ invariant $n^\text{int}_{12}$ is robust under the attachment of lower dimensional phases at the hinges or surfaces of a 3D sample.
Because the parity of $n_{12}/2$ matches that of the total number of hinge modes, $n_{12}/2 \, (\text{mod }2)$ is $\mathbb{Z}_2$ valued and captures the usual topological invariant for the $C_{4z}T$ invariant HOTI~\cite{schindler2018higher}.

We verify $n^\text{int}_{12}\in 2\mathbb{Z}$ by numerically computing the surface magnetizations and hinge currents for the intrinsic HOTI defined in Eq.~\eqref{eq:c4t_ham}. To obtain $K_1$, we follow the method outlined in Sec.~\ref{sec:Surface magnetization on the slab geometry} to calculate the layer-resolved magnetization $M_1(x)$ in a slab geometry with $10$ layers along the $\hat{x}$-direction, and then sum over the contributions localized to the surface normal to $+\hat{x}$. $K_2$ can be obtained in a similar manner. For $J_3$, we work in a geometry where the $\hat{z}$-direction is periodic and the other directions are open, with 80 unit cells in each direction, and sum over the contributions near the hinge shared by the $+\hat{x}$- and $+\hat{y}$-normal surfaces. 

\begin{figure}
\centering
\includegraphics[width=\linewidth]{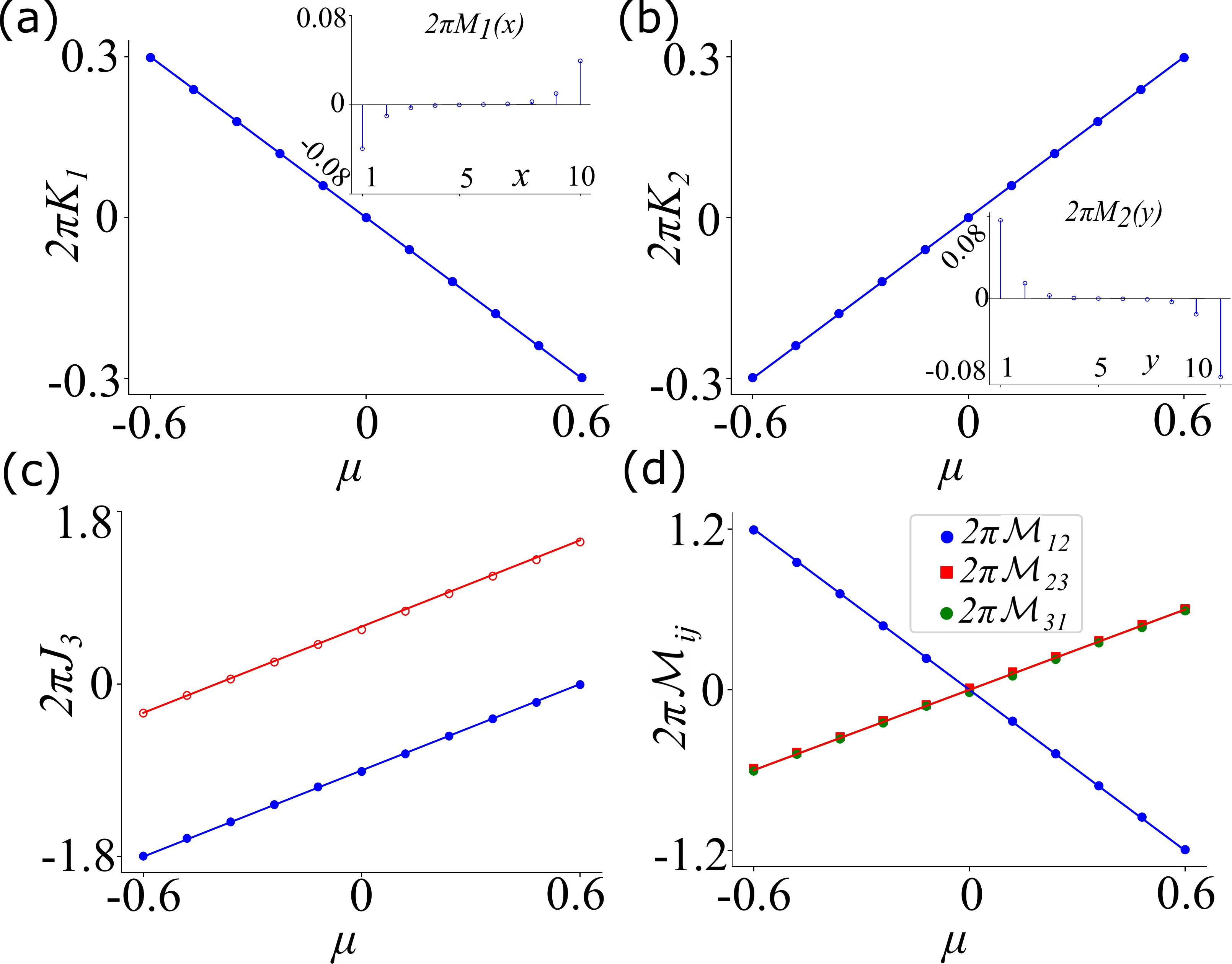}
 \caption{
Numerical calculations for the intrinsic HOTI in the topological phase. 
 (a) The surface magnetization $K_1$ as a function of the chemical potential $\mu$, calculated in a slab geometry with $10$ layers. 
 The line is a guide for the eyes showing that the slope of $K_1$ is approximately $-1/2$ when $\mu$ is within the gap of the system.
 In the inset, we show the layer resolved magnetization $M_1(x)$ for $\mu=0.6$. 
 (b)
 The surface magnetization $K_2$ as a function of the chemical potential $\mu$, calculated in a slab geometry with $10$ layers. 
 The slope of the $K_2$ is approximately $+1/2$. 
 In the inset, we show the layer resolved magnetization $M_2(y)$ for $\mu=0.6$. 
 (c)
 The hinge current $J_3$ as a function of the chemical potential $\mu$. The blue circles indicate the full hinge current and the red circles show the contribution from only the gapless hinge states. Both of the lines have slopes approximately $+1$. 
 (d)
 The bulk MQM as a function of $\mu$. The blue, red and green markers correspond to components $\mathcal{M}_{12}$, $\mathcal{M}_{23},$ and $\mathcal{M}_{31}$ respectively. The slope of $\mathcal{M}_{12}$ is $-2$ while the slopes for the other two components are $+1$. 
 }
 \label{fig:3}
\end{figure}

Explicitly, in Fig.~\ref{fig:3}, we consider numerics for the intrinsic HOTI in the topological phase with $M=2$ and $t=\Delta_1 = \Delta_2 = 1$ in Eq.~\eqref{eq:c4t_ham}. 
In Fig.~\ref{fig:3}(a-b), we show the surface magnetizations $K^\text{int}_1$ and $K^\text{int}_2$ as functions of the chemical potential. We see that their slopes are exactly opposite to each other with magnitude $1/2$, as expected from Eq.~\eqref{eq:K_1_K_2_C4zT}.
In the insets, we show the two layer-resolved magnetizations in their respective slab geometries and confirm that they are indeed localized near the surfaces, yielding well-defined surface magnetizations. 
From the classical relations between surface magnetizations and hinge currents in Eq.~\eqref{eq:hinge_current_decoration}, one may naively expect the hinge current at the intersection of the $+\hat{x}$ and $+\hat{y}$ surfaces to decrease as $\mu$ is increased. However, this contrasts with what we find in Fig.~\ref{fig:2}(c), where the full hinge current, shown as blue circles, in fact increases with the chemical potential with slope $+1$. This mismatch arises precisely because of the bulk MQM defined in Eq.~\eqref{eq:MQM_as_J_K}, which manifests as an additional \emph{bulk} contribution to the hinge current. 
The MQM component $\mathcal{M}_{12}$ is shown in Fig.~\ref{fig:3}(d), and its slope $n_{12}$ is $-2$, an even integer, as expected. In the trivial phase (not pictured), all surface magnetizations and hinge currents vanish, indicating zero bulk MQM.

Before moving on, there is a subtlety that we must address. Although we have defined $\mathcal{M}_{ij}$ as the net number of hinge modes arising from the bulk, it cannot be extracted by naively calculating the contribution of the bulk states to the hinge current.
In Fig.~\ref{fig:3}(c), the red circles correspond to the contribution to the hinge current arising purely from the low-energy hinge states, which is obtained by using the Lanczos method to isolate the hinge states in the band structure.
The difference between the current carried by the hinge states and the full hinge current gives the ``naive'' bulk contribution to the hinge current, which turns out to be independent of the chemical potential (c.f. Fig.~\ref{fig:3}(c)). This is in sharp contrast to the fact that the MQM has a nonzero derivative with respect to the chemical potential. Such a discrepancy can be attributed to fact that, much like the magnetic dipole moment, the MQM has two distinct contributions.
For the dipole case, it was shown using the Wannier representation that the 2D magnetization consists of a local contribution, which is $\mu$-independent, and an itinerant contribution, which is proportional to $\mu$ with proportionality constant equal to the Chern number~\cite{thonhauser2005orbital,Ceresoli2006}. 
By analogy, the bulk contribution to the hinge current uncovered in Fig.~\ref{fig:3}(c) can be interpreted as part of the local contribution to the MQM. 
It is unclear at the present moment whether a similar Wannier representation can be employed to distinguish the local and itinerant contributions to the MQM, and we leave this problem to future work.

\begin{figure*}
\centering
\includegraphics[width=\linewidth]{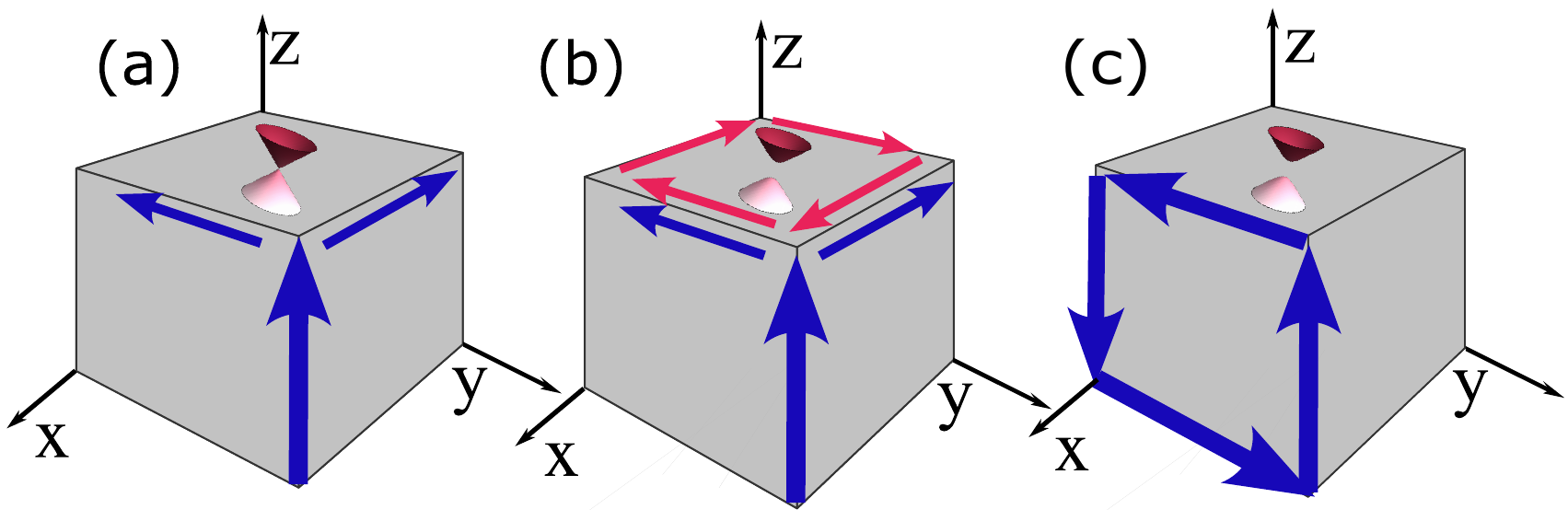}
\caption{(a) When $C_{4z}T$ symmetry is preserved the $z$-surface is gapless and the current splits evenly at the corner because of the symmetry. (b) When the symmetry is broken on the surface, the surface Dirac cone is generically gapped and a circulating current is induced that cancels $J_1$ and enhances $J_2$. (c) Assuming there is only one chiral hinge mode along each $\hat{z}$-parallel hinge, the resulting net hinge current circulates around the $\hat{x}$-normal surface.}
\label{fig:4}
\end{figure*}

We have thus far limited our focus to a single component of the bulk MQM, $\mathcal{M}_{12}$, which is unambiguously defined because both the $\hat{x}$- and $\hat{y}$-normal surfaces are gapped. In order to calculate $\mathcal{M}_{23}$ and $\mathcal{M}_{31}$, one would need to calculate the surface magnetizations for the $\hat{z}$-normal surfaces, which however host a gapless Dirac cone protected by the $C_{4z}T$ symmetry. The absence of a surface gap presents an obstacle for applying the local marker in Eq.~\eqref{eq:MQM_real_space_1} to calculate the layer-resolved magnetization on the $\hat{z}$-normal surfaces~\cite{bianco2013orbital}.
To bypass this difficulty, we take advantage of the fact that the bulk MQM is invariant under arbitrary surface deformations as long as they carry zero net magnetic dipole moment.
In particular, we consider a 2D decoration that infinitesimally breaks $C_{4z}T$ on the $\hat{z}$-normal surfaces such that the Dirac cones are gapped out~\cite{Essin2009Magnetoelectric}. In order to preserve the bulk MQM, the surface mass term must respect the three two-fold rotation symmetries $C_{2x,2y,2z}$. Using the representations given in Eq.~\eqref{eq:2_fold_symmetries}, we can identify 
\begin{align}
\Delta_{C_4T} = \delta \tau_2 \sigma_0,
\end{align}
where the constant $\delta\neq 0$, as the desired mass term to render both $K_3$ and $J_{1,2}$ well-defined.

Upon adding the mass term $\Delta_{C_4T}$ on the $+\hat{z}$ surface (and $-\Delta_{C_4T}$ on the $-\hat{z}$ surface), the surface Dirac cone is gapped, and the surface magnetization $K_3$ can be unambiguously calculated. Once again, we can use the relationship to the bulk magnetoelectric polarizability to write
\begin{align}
\label{eq:K_3_C4zT}
    2\pi \frac{dK_3^\text{int}}{d\mu} \in \begin{cases}
  \mathbb{Z}+\frac{1}{2} & \text{topological}\\   
  \mathbb{Z} & \text{trivial}
\end{cases},
\end{align}
which is identical to Eq.~\eqref{eq:K_1_K_2_C4zT}, but well-defined only after the $\hat{z}$-normal surfaces are gapped out. Importantly, Eqs.~\eqref{eq:K_1_K_2_C4zT} and \eqref{eq:K_3_C4zT} continue to hold because the surface decoration does not break $C_{4z}T$ symmetry in the bulk.

We can now repeat the same arguments we applied to $n_{12}$ to show that $n_{23}$ (or equivalently $n_{31}$) is an integer-valued invariant. Inside the surface gap, the contribution of the hinge current to $n_{23}$ must again be an integer counting the net number of chiral hinge modes (c.f. Eq.~
\eqref{eq:curr_deriv}). On the other hand, we have just shown that the contribution of the surface magnetizations to $n_{23}$ are all well-defined and half-integer (integer) in the topological (trivial) phase. The same considerations apply to $n_{31}$, and the definition of bulk MQM therefore yields
\begin{equation}
    \label{eq:n23n31}
    n_{23}, n_{31} \in \mathbb{Z}.
\end{equation}
Note that $C_{4z}T$ imposes a stronger condition on $n_{12}$ than the other components of bulk MQM: in the former case, surface magnetizations on adjoining $x$- and $y$-normal surfaces must also be equal and opposite, allowing only even values of $n_{12}$. On the other hand, we numerically confirm that $n_{23} = n_{31} = 1$ in Fig.~\ref{fig:3}(d), consistent with the tracelessness of the MQM tensor.

In addition, while changing the sign of the mass term can give rise to qualitatively different configurations of hinge current, it does not change the bulk MQM.
In Fig.~\ref{fig:4}(a), we show the hinge current configuration for the intrinsic HOTI before applying the surface mass term, where only three hinge currents are shown for clarity. Current conservation (c.f. Eq.~\eqref{eq:Conservation_currents}) and $C_{4z}T$ symmetry imply $-J_1=-J_2=J_3/2$. After the surface Dirac cones are gapped, the surfaces acquire a half-integer (integer) anomalous Hall conductivity in the topological (trivial) phase. 
Induced currents therefore circulate around the $\hat{z}$-normal surfaces, as shown in Fig.~\ref{fig:4}(b). The resulting pattern of hinge currents is akin to gluing one-half of a Chern insulator on the top and bottom surfaces, which (taking $\delta<0$ for concreteness) cancels $J_1$ and enhances $J_2$, as shown in Fig.~\ref{fig:4}(c).

\begin{figure}
\centering
\includegraphics[width=\linewidth]{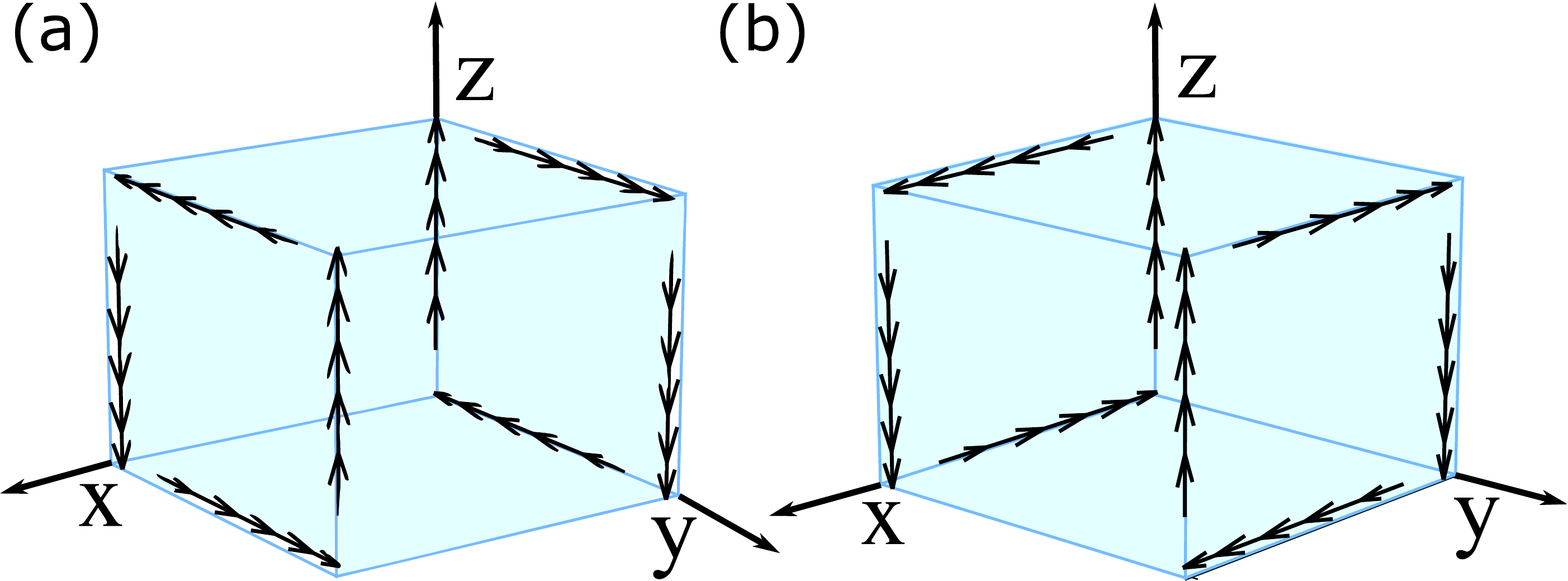}
\caption{Configuration of currents in the intrinsic $C_{4z}T$-symmetric HOTI with fully open boundary conditions. Depending on the sign of $\delta$ in the surface symmetry breaking term $\Delta_{C_{4z}T} =  \delta  \, \tau_2 \sigma_0$, we either have circulating current on (a) the $\hat{x}$-normal surfaces for $\delta<0$, or (b) the $\hat{y}$-normal surfaces for $\delta>0$. Here we consider a $6 \times 6 \times 6$ system with $\delta=0.3$, $\mu=0.2$.}
 \label{fig:5}
 \end{figure}

Changing the sign of $\delta$ reverses the Hall response on the $z$-normal surfaces, and therefore the direction of the resulting current. It follows that $\delta<0$ and $\delta>0$ yield a net circulating current on the $x$-normal and $y$-normal surfaces, respectively. We confirm this numerically by calculating the hinge currents in a fully open geometry. We indeed find that changing the sign of the mass term on the top and bottom layers indeed produces the expected current configurations as shown in Fig.~\ref{fig:5}.

To show our results are self-consistent, we now demonstrate that although the hinge current configurations are qualitatively different for $\delta<0$ and $\delta>0$,
the sign of the surface mass does not affect the topological invariants $n_{23}$ and $n_{31}$.
Recalling that
\begin{align}
    n_{23}=2\pi\frac{dM_{23}}{d\mu} = 2\pi\frac{d}{d\mu}\left(K_2-K_3-J_1\right),
\end{align}
and that $K_2$ is not affected by the surface mass term, we focus on the last two quantities. 
Comparing Figs.~\ref{fig:5}(a) and~\ref{fig:5}(b), we note that switching from $\delta<0$ to $\delta>0$ is equivalent to gluing a Chern insulator with Chern number $+1$ on the $z$-normal surfaces. As this has a positive-oriented $K_3$, we have
\begin{align}
    2\pi\frac{d}{d\mu}(K^\text{int}_3(\delta>0)-K^\text{int}_3(\delta<0)) = 1,
\end{align}
whereas
\begin{align}
    2\pi\frac{d}{d\mu}(J^\text{int}_1(\delta>0)-J^\text{int}_1(\delta<0)) = -1,
\end{align}
because $J^\text{int}_1(\delta>0)$ runs in the $-\hat{x}$ direction. Combining the two relations, we conclude that $2\pi\frac{d}{d\mu}(K^\text{int}_3+J^\text{int}_1)$, and hence $n^\text{int}_{23}$, is unaffected by the sign of the surface mass term. 

In summary, we have shown that the derivative of the bulk MQM, $n_{ij}$, is a topological invariant for the intrinsic HOTI that counts the net number of chiral hinge modes arising from the bulk. 
On $z$-parallel hinges, $C_{4z}T$ further requires the invariant $n_{12}$ to be an \emph{even} integer, and ${n_{12}}/{2}\, (\text{mod } 2)$ reduces to the standard $\mathbb{Z}_2$ invariant introduced in Ref. \onlinecite{schindler2018higher} to classify the chiral HOTI phase. The bulk MQM can give a finer classification than the $\mathbb{Z}_2$ invariant since it encodes the total number of hinge modes and not only their parity.
On the other hinges, the gapless $\hat{z}$-normal surfaces appear to present an obstacle to calculating the invariants $n_{23}$ and $n_{31}$. Nevertheless, the addition of a surface-localized $C_{4z}T$ symmetry-breaking term does not affect the bulk MQM and allows for its calculation in terms of boundary responses. 

In the case of the specific $C_{4z}T$-symmetric chiral HOTI, we have seen that the bulk MQM invariant agrees with the existing topological classification. However, our proof that $n_{12}$ is an even integer holds for general systems with $C_{4z}T$ symmetry and gapped surfaces. The MQM can then also be used to classify more exotic higher-order phases beyond the free-fermion picture of Ref.~\onlinecite{schindler2018higher}.
Indeed, a similar pattern of chiral hinge modes is found in the 3D strongly-interacting higher-order topological superconductor first described in Ref.~\onlinecite{May-Mann2022}. In this system, there are two chiral fermion modes (four chiral Majorana modes) on the hinges. Because the system is superconducting, these modes do not carry charge current, yet a thermal analog of our bulk MQM can be considered. Interestingly, the system has hinge modes, but $\mathbb{Z}_2$ subsystem symmetries pin the analog of the bulk $\theta$-angle and the analog of surface magnetizations to zero. The resulting MQM response is then very similar to that of the $C_{4z}T$-invariant intrinsic HOTI, with two net chiral hinge modes arising from the bulk:
\begin{equation}
    \label{eq:interacting_inv}
    n_{12} = - 2\pi  \frac{d J_3}{d\mu} = -2.
\end{equation}
Unlike the intrinsic HOTI, only the hinge current contributes to the bulk MQM, but the resulting invariant is the same.
The (thermal) MQM of the interacting system then obeys our criteria for $n_{12}\in 2\mathbb{Z}$. This relationship suggests that the MQM may also shed light on other strongly-interacting systems with exotic symmetries, a question that we leave to future work.

\subsection{MQM in an extrinsic HOTI}
The arguments of the previous section show that the MQM can be used to distinguish the trivial and non-trivial phases of an intrinsic HOTI in the presence of a symmetry like $C_{4z}T$.
There also exists a large class of extrinsic (e.g., boundary-obstructed) higher-order phases whose boundary signatures are strongly dependent on the choice of termination. 
Unlike the $C_{4z}T$ invariant HOTI, 
the hinge currents of these phases can be completely removed by gluing lower dimensional phases to their surfaces.
It is therefore critical to define topological invariants to distinguish such phases from completely trivial insulators. In this section, we consider an example of a 3D extrinsic HOTI and show that the bulk MQM can be such a topological invariant. 

To create a model of a 3D HOTI that, unlike the $C_{4}T$ HOTI, has no intrinsic bulk obstruction, consider the boundary obstructed, pumped-quadrupole model from Ref. \onlinecite{benalcazar2017B}: 
\begin{align}
\label{eq:extrinsic_ham}
H^\text{ext}({\bf k})  = &- \sin k_x \tau_2 \sigma_3+ (-2+m_x+\cos k_x+\cos k_z)\tau_1 \notag\\
-& \sin k_y\tau_2 \sigma_1 - (-2+m_y+\cos k_y+\cos k_z)\tau_2 \sigma_2 \notag\\
+& \sin k_z\tau_3.
\end{align}
We note that the last term breaks $T$ and mirror symmetries, but preserves the three twofold rotation symmetries along the three axes, with representations $C_{2x}=\tau_1\sigma_1$, $C_{2y} =\tau_1\sigma_3$, and $C_{2z}=\tau_0\sigma_2$. As we saw earlier, these symmetries pin the bulk magnetic dipole moments to zero, allowing for a well-defined MQM.  We now show that for this class of HOTIs, the derivatives of the MQM are $\mathbb{Z}$-classified:
\begin{align}
    n^\text{ext}_{ij}\in \mathbb{Z}.
\end{align}

In contrast to the intrinsic HOTI considered in Sec.~\ref{sec:MQM in higher order topological phases}, the surfaces of $H_\text{ext}$ are all gapped because the surface Dirac cones, if any, are not protected by the twofold rotation symmetries. It follows that all boundary quantities can be calculated unambiguously.
Making use of the methods in Sec.~\ref{sec:Surface magnetization on the slab geometry} to calculate the layer resolved magnetizations directly (without the need to apply surface symmetry-breaking masses), we find that the resultant surface magnetizations are again related to the bulk magnetoelectric polarizability as
\begin{align}
\label{eq:Ks_ext}
    2\pi \frac{dK_1}{d\mu} = 2\pi \frac{dK_2}{d\mu} = 2\pi \frac{dK_3}{d\mu} = \theta \, (\text{mod  } 1).
\end{align}
Unlike the $C_\text{4z}T$ symmetry present in the intrinsic case, none of the twofold rotation symmetries $C_{2x,2y,2z}$ can quantize the bulk $\theta$-angle. Nevertheless, from Eq.~\eqref{eq:Ks_ext} we can conclude that the differences of the derivatives are all integer valued:
\begin{equation}
\begin{aligned}
    &2\pi \frac{d(K_1-K_2)^\text{ext}}{d\mu} = 2\pi \frac{d(K_2-K_3)^\text{ext}}{d\mu}=2\pi \frac{d(K_3-K_1)^\text{ext}}{d\mu} \\
    &= 0 \, (\text{mod  } 1).
\end{aligned}
\end{equation}
Because all the surfaces are gapped, the arguments in Sec.~\ref{sec:MQM in HOTIs} still apply to the hinge currents, whose derivatives are then also integer valued:
\begin{align}
    2\pi \frac{dJ_1^\text{ext}}{d\mu},\, 2\pi \frac{dJ^\text{ext}_2}{d\mu},\, 2\pi \frac{dJ_3^\text{ext}}{d\mu} \in \mathbb{Z}.
\end{align}
Upon combining the two relations, we arrive at the desired result, $n_{ij}^\text{ext}\in\mathbb{Z}$. 

\begin{figure}
\centering
  \includegraphics[width=\linewidth]{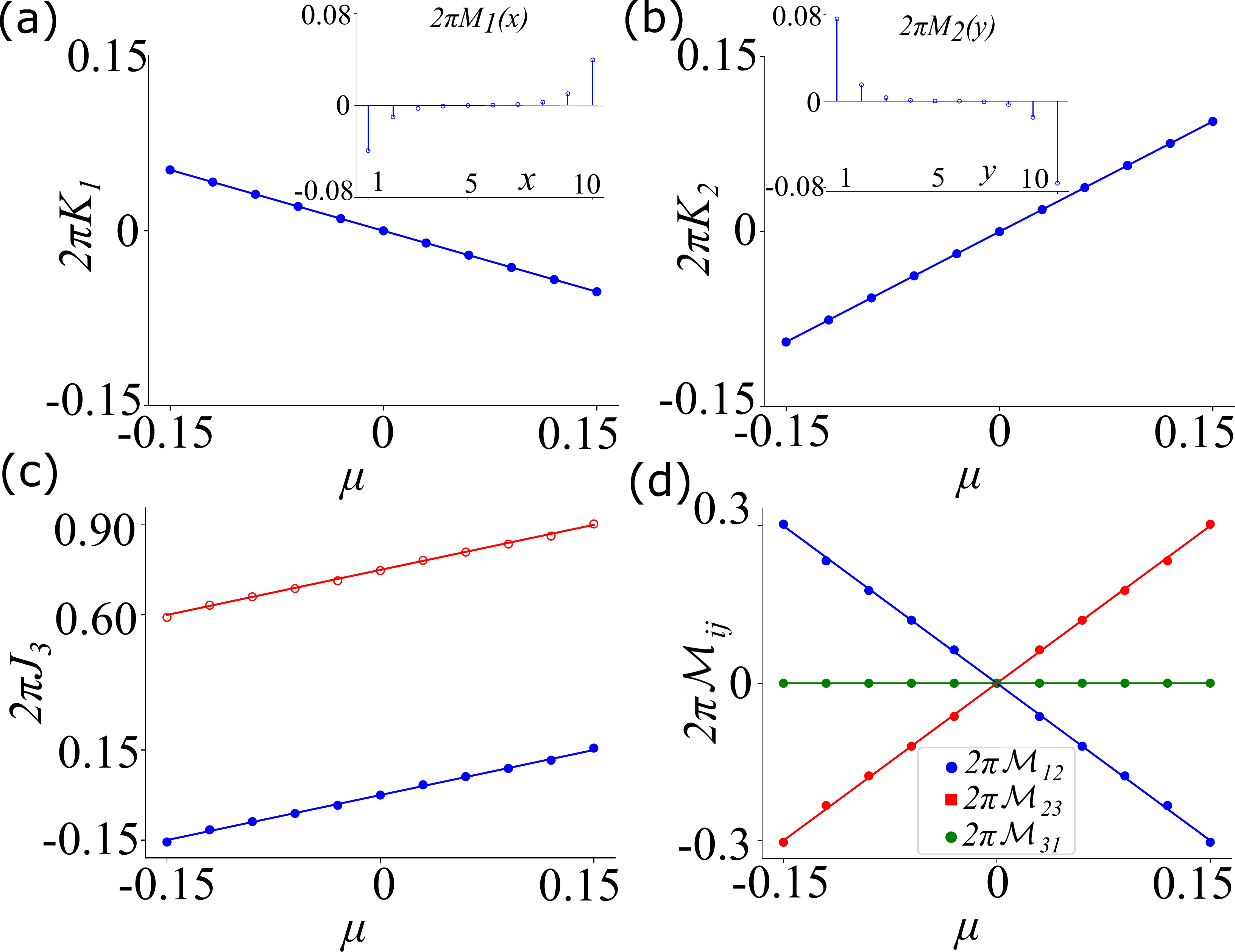}
 \caption{
 Numerical calculations for the extrinsic HOTI. 
 (a) The surface magnetization $K_1$ as a function of the chemical potential $\mu$, calculated in a slab geometry with $10$ layers. 
 The line is a guide for the eyes showing that the slope of $K_1$, in contrast to Fig.~\ref{fig:3}(a), is not integer valued.
 In the inset, we show the layer resolved magnetization $M_1(x)$ for $\mu=0.15$. 
 (b)
 The surface magnetization $K_2$ as a function of the chemical potential $\mu$, calculated in a slab geometry with $10$ layers. 
 The slope of the $K_2$ is again not integer valued, in contrast to the intrinsic case shown in Fig.~\ref{fig:3}(b). 
 In the inset, we show the layer resolved magnetization $M_2(y)$ for $\mu=0.15$. 
 (c)
 The hinge current $J_3$ as a function of the chemical potential $\mu$. The blue circles indicate the full hinge current and the red circles show the contribution from only the gapless hinge states. Both of the lines have slopes approximately $+1$, matching the the intrinsic case in Fig.~\ref{fig:3}(c).
 (d)
 The bulk MQM as a function of $\mu$. The blue, red and green markers correspond to components $\mathcal{M}_{12}$, $\mathcal{M}_{23},$ and $\mathcal{M}_{31}$ respectively. $\mathcal{M}_{12}$ and $\mathcal{M}_{23}$ have slopes $-2$ and $+2$ respectively, while $\mathcal{M}_{31}$ has a slope of zero. 
 }
 \label{fig:6}
 \end{figure}

The quantization of the invariants can also be confirmed numerically. In Fig.~\ref{fig:6}(a), we show the surface magnetizations $K_1^\text{ext}$ and $K^\text{ext}_2$ as functions of the chemical potential with $(m_x,m_y)=(0.2, 0.5)$ in Eq.~\eqref{eq:extrinsic_ham}. The corresponding layer-resolved magnetizations are shown in the inset. 
The layer-resolved magnetization $M_3(z)$ and $K^\text{ext}_3$ share a similar pattern and hence are not shown for clarity. 
Because the derivatives $2\pi\frac{dK^\text{ext}_i}{d\mu}$ are not quantized by any symmetries, their values depend on the parameters of the system.
For example, the slopes in Fig.~\ref{fig:6}(a-b) are found to be $-0.35$ and $0.64$, respectively.
This is in sharp contrast to Fig.~\ref{fig:3}(a-b) for the intrinsic HOTI, where the slopes are half-integers. Nevertheless, their difference is indeed quantized, as expected from Eq.~\eqref{eq:Ks_ext}. 

In Fig.~\ref{fig:6}(c) we show the full hinge current and the low-energy hinge contribution in blue and red circles, respectively, both of them having slope $+1$. As discussed in Sec.~\ref{sec:MQM in HOTIs}, their difference, which is independent of chemical potential, can be regarded as the local contribution to $\mathcal{M}_{12}$. 
The bulk MQM as a function of chemical potential is shown in Fig.~\ref{fig:6}(d), where $2\pi\mathcal{M}_{31}$ remains constant as $\mu$ increases, whereas $2\pi\mathcal{M}_{12}$ and $2\pi\mathcal{M}_{23}$ have slopes $-2$ and $+2$, respectively, in accordance with the tracelessness of the full MQM tensor. We also mention that in the trivial phase of the HOTI (not pictured), there are no hinge currents or surface magnetizations, yielding no bulk MQM.

Unlike the intrinsic case, we can completely eliminate the hinge modes of the extrinsic HOTI by symmetrically gluing two Chern insulators to the $y$-normal surfaces, but the bulk MQM does not change. A similar process applied to a trivial 3D bulk \emph{generates} hinge currents, thus switching the response of the ``topological'' and ``trivial'' phases. Nevertheless, the topological distinction between the phases remains, as the former phase maintains ${n_{23} = -n_{12} = 2}$, while the latter has zero MQM. Although the hinge currents alone cannot distinguish the two phases because they are not invariant under such gluing procedures, the bulk MQM is robust and quantifies the boundary-obstruction between the extrinsic HOTI and a completely trivial system. In passing, we note that the same phenomenology can be found in other extrinsic HOTIs. For example, a stack of 2D bilayers with opposite Chern numbers, in the phase where the layers are strongly coupled between unit cells, has unpaired Chern insulators left over on the top and bottom surfaces. Such a system has a pattern of hinge currents and surface magnetizations that yields the same bulk MQM as the extrinsic HOTI considered above, and is itself also a boundary-obstructed HOTI.

\subsection{MQM in a trivial HOTI}
\label{sec:MQM for trivial HOTI}

In the previous two sections, we have examined two examples with nonzero bulk MQM, both exhibiting surface magnetizations and hinge currents.
Here we consider an example with similar boundary observables but zero MQM in the bulk. Exploiting the invariance of the MQM under gluing lower-dimensional phases, we begin with a completely trivial bulk and attach 2D Chern insulators to the top and bottom surfaces. In this manner, we expect that the bulk MQM remains zero, but the 2D decorations generate a surface magnetization and hinge currents. 

Explicitly, for the trivial 3D bulk, we consider the Hamiltonian
\begin{equation}
    \label{eq:trivial}
    H^\text{triv} = m \, \tau_2 \sigma_3,
\end{equation}
which for $m\neq 0$ is both gapped, $C_{2}$-symmetric, and time-reversal invariant if $\tau$ is orbital and $\sigma$ is spin. Consequently, $H^\text{triv}$ cannot support hinge currents or surface magnetizations, and its bulk MQM is identically zero. On the top and bottom surfaces, we attach a two-dimensional Chern insulator given by 
\begin{align}
    \label{eq:CI}
    \begin{split}
    H^{\text{surf}} = &\sin{k_x} \tau_0\sigma_1 + \sin{k_y} \tau_0\sigma_2 \\
    & + (2 + M - \cos{k_x} - \cos{k_y}) \tau_0\sigma_3.
    \end{split}
\end{align}
When $M=-3$, this model has Chern number $C=-2$, indicating that two chiral hinge modes circulate around its edges in an open geometry. Attaching $H^\text{surf}$ to the top surface, and $-H^\text{surf}$ to the bottom surface breaks time-reversal while preserving the $C_2$ symmetries. The MQM, which is thus still well-defined, cannot change under such a gluing procedure and remains trivial.

\begin{figure}
\centering
  \includegraphics[width=\linewidth]{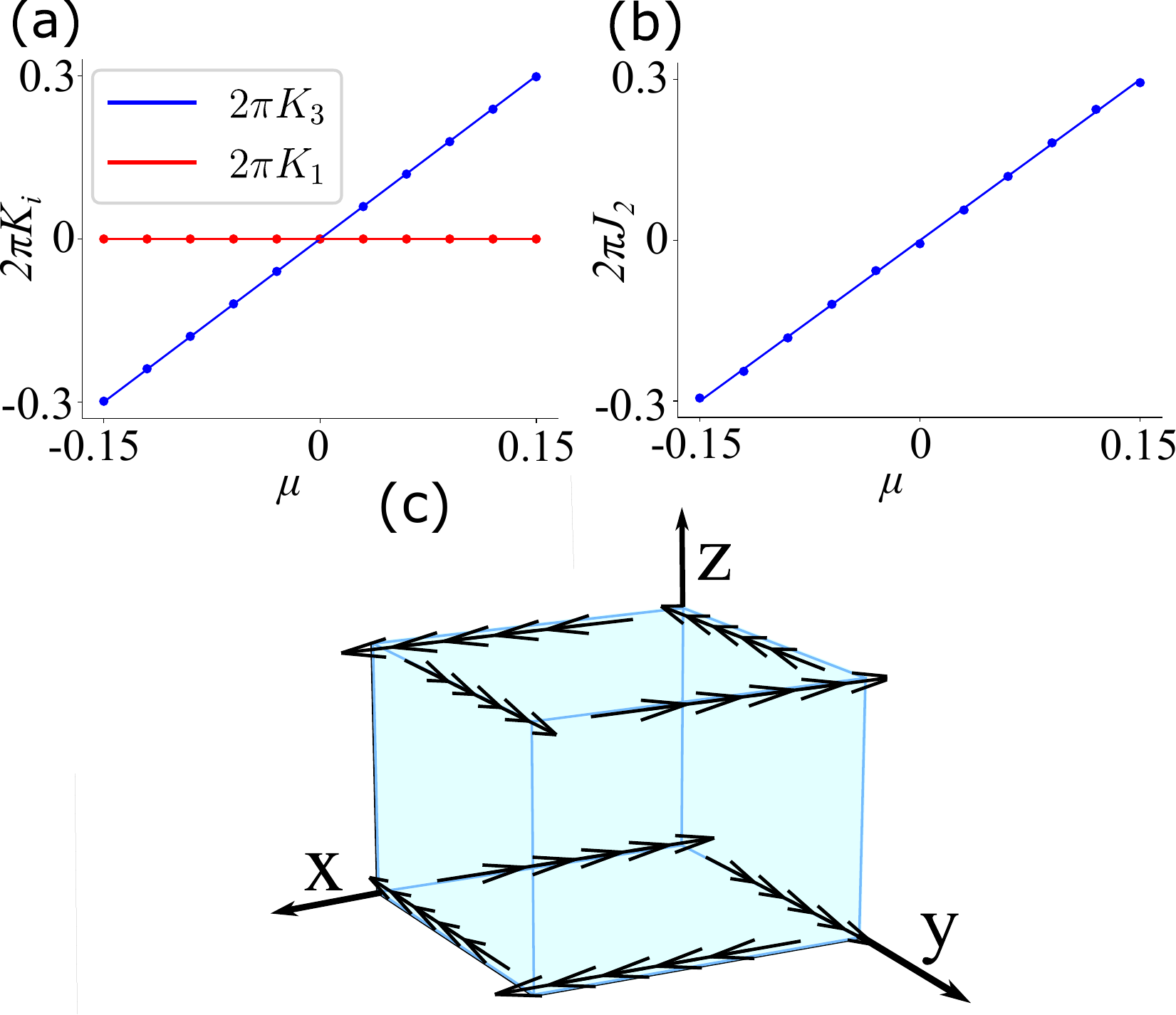}
 \caption{
 Calculations for a trivial bulk insulator with 2D Chern insulators with $C=-2$ glued to the top and bottom surfaces.
 (a) The surface magnetizations $K_3$ and $K_1$ as a function of the chemical potential $\mu$, calculated in a slab geometry with $10$ layers. 
 The line indicates that the slope of $K_3$ is $1/\pi$, while the slope of $K_1$ is zero.
 (b)
 The hinge current $J_2$ as a function of the chemical potential $\mu$, calculated in a geometry with $y$ periodic, and $x,z$ open where $N_x=N_z=20$. 
 The line shows that the slope of the hinge current matches that of the surface magnetization, indicating that $\mathcal{M}_{31} = 0$. The other components of the bulk MQM (not shown) are also zero.
 (c)
 The hinge currents $J_1$ and $J_2$ in a fully open geometry with $N_x = N_y = N_z = 6$. Here the currents arise entirely because of the surface magnetization.}
 \label{fig:7}
 \end{figure}

We confirm this numerically by calculating the surface magnetizations and hinge currents of the Chern insulators with $M=-3$ attached to the $z$-normal surfaces of a trivial bulk with $m=1$. In Fig.~\ref{fig:7}(a) we show the dependence of $K_3$, the only nontrivial surface magnetization, on chemical potential. The slope is proportional to the Chern number, as in a two-dimensional system~\cite{Ceresoli2006}, and is equal to the slope of the hinge current $J_2$, shown in Fig.~\ref{fig:7}(b). It therefore follows that $\mathcal{M}_{31} = K_3 - K_1 - J_2 = 0$, and the other components of the bulk MQM similarly vanish. In Fig.~\ref{fig:7}(c), we show the configuration of hinge currents in a fully open geometry. Although the pattern of circulating currents is almost identical to that of the intrinsic HOTI shown in Fig.~\ref{fig:5}, here they arise entirely from the surface magnetization $K_3$ as opposed to a bulk MQM. 

\section{Symmetry constraints on the MQM}\label{sec:symmetry}
To better understand the various MQM responses of the HOTI systems that we have examined, we now study how the MQM transforms under more general symmetries of an orthorhombic lattice. As before, the inherent twofold rotation symmetries $C_{2x}$, $C_{2y}$, and $C_{2z}$ of these lattices ensure that the bulk magnetic dipole moment vanishes, and hence the MQM is well-defined. 
Based on our definition of the MQM in Eq.~\eqref{eq:MQM_as_J_K}, we can determine its symmetry properties from those of the $J_i$ and $K_i$. In what follows, we systematically derive symmetry constraints on the various components of $\mathcal{M}_{ij}$ and further apply them to our HOTI invariants $n_{ij}$. Because of the twofold rotation symmetries, it is enough to consider the boundary responses near a single corner, as in Fig.~\ref{fig:1}(c), and determine the sign of the transformed responses using our sign conventions. Recall that positive hinge currents are defined in Fig.~\ref{fig:1}(a), and positive surface magnetizations point in the outward normal direction.
\subsection{Constraints from $C_{4}$ and $C_{4}T$ rotation symmetries}
We first consider fourfold rotation symmetry around the $z$-axis, {$C_{4z}: (x,y,z) \rightarrow (y, -x, z)$}. Under this symmetry, the surface responses transform as:
\begin{align}
    \label{eq:c4rotation_surf}
    \begin{split}
    (K_1, K_2, K_3) &\rightarrow (K_2, K_1, K_3), \\
    (J_1, J_2, J_3) &\rightarrow (-J_2, -J_1, -J_3).
    \end{split}
\end{align}
These transformations imply that the transformation of the bulk MQM is given by:
\begin{equation}
    \label{eq:c4rotation_mqm}
    C_{4z}: (\mathcal{M}_{12}, \mathcal{M}_{23}, \mathcal{M}_{31}) \rightarrow (-\mathcal{M}_{12}, -\mathcal{M}_{31}, -\mathcal{M}_{23}).
\end{equation}
Hence, $C_{4z}$ symmetry constrains $\mathcal{M}_{12} = 0$ and $\mathcal{M}_{23} = - \mathcal{M}_{31}$. Consequently, in the presence of two orthogonal fourfold rotation symmetries, the full MQM must vanish. None of the models we have considered have all three $C_4$ symmetries, but the ``trivial HOTI'', i.e. Chern insulators glued to the $z$-normal surfaces of a trivial 3D bulk, has $C_{4z}$ symmetry, requiring $\mathcal{M}_{12}=0,$ just as we determined in Sec.~\ref{sec:MQM for trivial HOTI}.

As both hinge currents and surface magnetizations change sign under time reversal $T$, the combination of $T$ with $C_4$ leads to additional symmetry constraints. In particular, the MQM transforms under the composite symmetry as 
\begin{equation}
    \label{eq:c4trotation_mqm}
    C_{4z}T: (\mathcal{M}_{12}, \mathcal{M}_{23}, \mathcal{M}_{31}) \rightarrow (\mathcal{M}_{12}, \mathcal{M}_{31}, \mathcal{M}_{23}).
\end{equation}
Thus, in a $C_{4z}T$-symmetric system, like the intrinsic HOTI of Sec.~\ref{sec:MQM in HOTIs}, $\mathcal{M}_{23} = \mathcal{M}_{31}$. Furthermore, the tracelessness of full MQM tensor requires the components of the bulk MQM to obey
\begin{equation}
    \label{eq:traceless_bulk_mqm}
    \mathcal{M}_{12} + \mathcal{M}_{23} + \mathcal{M}_{31} = 0.
\end{equation}
Using this additional constraint, we find that ${\mathcal{M}_{12} = -2 \mathcal{M}_{23}}$ in general $C_{4z}T$-symmetric systems. We also see that two or more orthogonal $C_4T$ symmetries are present, all components of the bulk MQM must vanish.

In the previous section, we showed that some phases exhibit a quantized invariant $n_{ij} = 2\pi \frac{d \mathcal{M}_{ij}}{d\mu}$. The constraint from $C_{4z}T$ immediately implies that $n_{12}$ is an even integer, as we showed. 

\subsection{Constraints of reflection symmetries and improper rotations}

We now turn to the effects of reflection symmetries $R_x$, $R_y$, and $R_z$. The three, twofold rotation symmetries are themselves the product of two mirror reflections (e.g., $C_{2z} = R_x R_y$, etc.). Hence, to keep all three $C_{2i}$ symmetries, the systems we consider can either have all three mirror symmetries, or none of them. 

Let us consider the case of a system having all three reflection symmetries. Because of the constraints already arising from the $C_2$ symmetries, any of the $R_i$ maps all the hinge currents in Fig.~\ref{fig:1}(a) to their negatives. Similarly, any of the mirror symmetries also maps all surface magnetizations to their negatives. To see this, let us concentrate on the action of $R_z: (x,y,z)\rightarrow(x,y,-z)$. Heuristically, the magnetic dipole moment is $\bf{M} \sim \bf{r} \times \bf{v}$, and thus $M_1$ and $M_2$ manifestly change sign under an $R_z$ transformation. As this symmetry does not interchange any $x$- and $y$-normal surfaces, this in turn implies that $K_1$ and $K_2$ change sign. On the other hand, the $M_3$ component of the layer-resolved magnetization does not change sign under $R_z$, yet the symmetry maps one $z$-normal surface to the other, exchanging the direction of the outward normal and thus the sign of $K_3$. Combining the transformation properties of the currents and surface magnetizations, we arrive at
\begin{equation}
    \label{eq:reflection_mqm}
    R_i: (\mathcal{M}_{12}, \mathcal{M}_{23}, \mathcal{M}_{31}) \rightarrow (-\mathcal{M}_{12}, -\mathcal{M}_{23}, -\mathcal{M}_{31}).
\end{equation}
Hence, in orthorhombic crystals, any additional reflection symmetry is enough to pin all the bulk MQM components to zero. Like time-reversal, reflection symmetries must therefore be broken in orthorhombic HOTI models with nontrivial MQM.

 Now that we understand the action of both reflections and $C_{4}$ rotations We can also study the effect of improper rotations, i.e., the product of rotations and reflections. Let us consider an $S_{4z,i}=C_{4z}R_i$ . Combining Eqs.~\eqref{eq:c4rotation_mqm} and \eqref{eq:reflection_mqm}, the net effect of a $S_{4z,i}$ transformation is identical to that of a $C_{4z}T$ transformation. In HOTIs with a quantized, integer $n_{ij}$, this implies that $n_{12}$ is even, and $n_{23} = n_{31}$. Again, if more than one such symmetry is present for orthogonal rotation axes, then the bulk MQM is identically zero.

\subsection{Constraints from $C_3$ rotation symmetry.}
Finally, let us consider ``cubic" $C_3$ rotations around an axis through  a corner of our system. Such operations permute the $x$-, $y$-, and $z$-normal surfaces, and  from Fig.~\ref{fig:1}{(c)} we see that the surface magnetizations and hinge currents transform as
\begin{align}
    \label{eq:c3rotation_surf}
    \begin{split}
    (K_1, K_2, K_3) &\rightarrow (K_2, K_3, K_1) \\
    (J_1, J_2, J_3) &\rightarrow (J_2, J_3, J_1).
    \end{split}
\end{align}
Hence, the MQM transforms as:
\begin{align}
    \label{eq:c3rotation_mqm}
    C_3: (\mathcal{M}_{12}, \mathcal{M}_{23}, \mathcal{M}_{31}) \rightarrow (\mathcal{M}_{23}, \mathcal{M}_{31}, \mathcal{M}_{12}).
\end{align}
Applying the traceless constraint, we immediately see that this $C_3$ rotation symmetry forbids any nonzero bulk MQM components. As we have seen, the bulk MQM is heavily constrained by symmetries and can be nonzero in orthorhombic systems only if the symmetry group does not contain reflections or $C_3$.

\section{Discussion}
\label{sec:Discussion}

\subsection{Relation to previous work}
We now comment on the relation between the current work and earlier, published results that calculate the MQM using an alternative method. In Refs.~\onlinecite{shitade2018theory,gao2018orbital}, the orbital MQM is studied using a Green function formalism and semiclassical dynamics, respectively, and the results seem to agree with each other. 
However, as we now explain, their formulas are not applicable to the higher order topological phases discussed here, and thus finding a bulk formula for the MQM (i.e., using periodic boundary conditions in all directions), at least for higher order topological phases, remains an open question.

Let us now briefly compare our results to this previous work. In Ref.~\onlinecite{shitade2018theory}, it was found that the derivative of the MQM with respect to chemical potential is equal to the cross-gap contribution to the linear magneto-electric susceptibility $\alpha^i_{\,  j}$ introduced  in Ref.~\onlinecite{essin2010orbital}. This contribution is gauge invariant and traceless $(\sum_{j=i}\alpha^i_{\,  j}=0)$, which are promising signs.
However, as shown in Ref.~\onlinecite{essin2010orbital}, $\alpha^i_{\,  j}=0$ identically if the bands satisfy the following ``degeneracy'' and ``reflection'' conditions:  
(i) all the valence bands are degenerate at $E_{\bf k}^v$ for a give{n} ${\bf k}$; (ii) all the conduction bands are degenerate at $E_{\bf k}^c$ for a given ${\bf k}$; (iii) $E_{\bf k}^c+E_{\bf k}^v$ is independent of ${\bf k}$.
It is for this reason that such cross-gap contributions to the magneto-electric susceptibility do not appear in certain models of 3DTIs~\cite{qi2008topological} and, importantly, HOTIs. In particular, the band structures of the HOTIs considered here also satisfy these conditions because their Bloch Hamiltonians are constructed using five anti-commuting $\Gamma$-matrices. Therefore, $\alpha^i_{\,  j}$ is always zero, and the formula from Ref.~\onlinecite{shitade2018theory,gao2018orbital} results in vanishing  MQM for our HOTI models. On the other hand, we know this cannot be true, as our results show clear macroscopic manifestations of a bulk MQM. 
Further investigation is therefore needed to understand the relationship between the results in Refs.~\onlinecite{shitade2018theory,gao2018orbital} and macroscopic manifestations of the diagonal components of the MQM, such as the hinge currents and surface magnetizations that are found in HOTIs.
For further comparison we note that  Ref.~\onlinecite{shitade2018theory}  proposes that the {anti-ferromagnetic} semiconductor BaMn$_2$As$_2$ can host nontrivial diagonal and off diagonal MQM, a claim supported with an effective model. 
{However, there seems to be no evidence that the model supports hinge currents}~\footnote{Private communication with Atsuo Shitade.} and thus
it is not clear what the boundary physical observable for such a MQM, and in particular its diagonal components, would be. Furthermore, in Ref.~\onlinecite{gao2018orbital}, a two band model with $H_{\bf k}=v'k_x+k_x\sigma_x+k_y\sigma_y+\Delta\sigma_z$ is used to demonstrate the presence of an \emph{off-diagonal} MQM component $M_{23}$. However, this is essentially {a} ${\bf k}\cdot{\bf p}$ expansion around a single node of a 2D DSM, which already has nonzero magnetization $M_3$ (because the nodes are offset in energy by the $v'k_x$ term~\cite{ramamurthy2015patterns}), and thus $M_{23}$ is in fact ill-defined.
{To see this explicitly, note that according to Eq.~\eqref{eq:shift_MQM}, in the presence of nonzero $M_3$, the response of the MQM to a shift of origin ${\bf r}\rightarrow{\bf r}+{\bf a}$ reads
\begin{eqnarray}\begin{aligned}
\label{eq:Shift_MQM_2D_DSM}
\delta M_{11} &=\delta M_{22} =\frac{2}{3}a_3M_3,\\
\delta M_{{33}} &=-\frac{4}{3}a_3M_3,\\
\delta M_{13} &=-2a_1M_3,\\
\delta M_{23} &=-2a_2M_3,\\
\delta M_{12} &=\delta M_{21} =\delta M_{31} =\delta M_{32} = 0.\\
\end{aligned}\end{eqnarray}
In this case, the diagonal components of the MQM are all origin-dependent, despite the fact that the trace remains zero. The only well-defined quantities are the four off-diagonal components at the end of Eq.~\eqref{eq:Shift_MQM_2D_DSM}, but $M_{23}$ is not one of them. 

In fact, the definition of off-diagonal MQM components is a very subtle issue for two dimensional electronic systems. Thermodynamically, the MQM can be defined as~\cite{raabbook,shitade2018theory,gao2018orbital}
\begin{eqnarray}\begin{aligned}
\label{eq:MQM_thermo_def}
M_{ij} = \frac{2}{3V}\frac{\partial \Omega}{\partial(\partial_iB_j)}\biggr\rvert_{T,\mu} ,
\end{aligned}\end{eqnarray}
where $\Omega= E-TS-\mu N$ is the grand canonical potential. 
Eq.~\eqref{eq:MQM_thermo_def} can be shown to be equivalent to the classical definition in Eq.~\eqref{eq:classical_MQM}. 
As pointed out in Ref.~\onlinecite{gao2018orbital}, since two dimensional electrons can couple only to $B_3$, it is only $M_{23}$ and $M_{13}$ that have clear physical meanings since, as per Eq.~\eqref{eq:MQM_thermo_def}, they are the in-plane gradients of the magnetization $M_3$. However, as shown in Eq.~\eqref{eq:Shift_MQM_2D_DSM}, these components suffer from origin-dependency unless $M_3=0$. On the other hand, well-defined components like $M_{12}$, have no clear physical meaning in two dimensions, as per Eq.~\eqref{eq:MQM_thermo_def}. 

Finally, we note that Eq.~\eqref{eq:Shift_MQM_2D_DSM} suggests that instead of requiring $M_{ij}$ to be a full well-defined rank-2 tensor, which is possible if and only if the magnetization is zero, we can relax this constraint slightly and consider well-defined \emph{individual} components of $M_{ij}$. 
This opens up an interesting avenue of research for exploring systems with well-defined off-diagonal MQM as in Ref. \onlinecite{winkler2020collinear}.
}

\subsection{Linear response theory for MQM}
\label{sec:Bulk formula for MQM}

An important goal is to understand the MQM purely from the bulk of an extended system. Our slab geometry approach provides a key step towards it by treating two out of three spatial directions with periodic boundary conditions. 
In light of this goal, one may wonder whether we could simply apply the perturbative formalism of Ref.~\onlinecite{shi2007quantum} to derive a bulk formula for the MQM.
Indeed, using Eq.~\eqref{eq:MQM_thermo_def} it seems straightforward to calculate $M_{ij}$ at zero temperature from linear response theory, provided that we consider a ${\bf B}$-field configuration with nonzero (preferably constant) gradient. For example, one could consider the $q_x, q_y\to 0$ limit of the vector potential
\begin{eqnarray}\begin{aligned}
\label{eq:A_config_2}
{\bf A} = B_{12}\frac{\sin q_xx}{q_x}\frac{\sin q_yy}{q_y} \hat{z} ,
\end{aligned}\end{eqnarray} for which the gradient of the ${\bf B}$-field reads ${\partial_1B_1= -\partial_2B_2 = B_{12}\cos q_xx \cos q_yy}$. 
We note that in the absence of either $q_x$ or $q_y$ dependence, the resulting ${\bf B}$-field would be constant with vanishing gradient~\cite{shi2007quantum}.

For the vector potential in Eq.~\eqref{eq:A_config_2}, at zero temperature, the variation of the free energy density is $\delta F(r) = B\cos q_xx\cos q_yy({M}_{11}-{M}_{22})$. From this result one immediately observes a complication: there is no way to separate the contributions from ${M}_{11}$ and $M_{22}$. This is very different from the calculation of the magnetization in Ref.~\onlinecite{shi2007quantum}. Indeed, this result suggests that the \emph{differences} between the $M_{ii}$, such as ${M}_{11}-{M}_{22}$,  are better observables than the ${M}_{ii}$ themselves. This matches exactly with our results since, if we recall Eq.~\eqref{eq:MQM_as_J_K}, ${M}_{11}-{M}_{22}$ is nothing but (twice) the bulk MQM $\mathcal{M}_{12}$, i.e., the difference represents the \emph{bulk} MQM. In perturbation theory we might expect turning on a vector potential to generate $\delta F\sim {\bf J}\cdot{\bf A}.$ Interestingly, our vector potential is in the $z$-direction, and the moment $\mathcal{M}_{12}$ receives contributions from the hinge currents parallel to $z,$ which matches the perturbation theory expectation. 

To determine the MQM one might then try to calculate the three hinge currents $J_i$ from perturbation theory and reconstruct $M_{ii}$ (or more precisely $\mathcal{M}_{ij}$ with the approach outlined in Sec.~\ref{sec:Classical definition of magnetic quadruple moments}. However, we are {only} cautiously optimistic that this is the right approach, as we expect that it will lead to the same result found by Refs.~\onlinecite{shitade2018theory,gao2018orbital}, which does not apply to higher order phases, as we have explained above. Thus, further research is needed to obtain a complete bulk formula of MQM that can be applied for higher order topological phases.

\section{Conclusion}
\label{sec:Conclusion}
In conclusion, we have studied the orbital magnetic quadrupole moment and applied our theory to three dimensional higher order topological phases. To this end, we derived a formula for calculating the surface localized magnetization, and showed that the bulk MQM is a linear combination of the surface magnetizations and hinge currents. 
The bulk MQM serves as a measure of the mismatch between the surface magnetizations and hinge currents, similar to how the electric quadrupole measures the difference between edge polarizations and corner charges. 
Although the MQM is defined as a combination of boundary observables, we have shown that it is a bulk quantity that is insensitive to surface decorations as long as the magnetization of the system remains zero, a prerequisite for well-defined MQM.

Focusing on two HOTIs with nontrivial MQMs, one intrinsic and one extrinsic, we have shown that the derivative of the MQM with respect to the chemical potential can serve as a topological invariant to distinguish the trivial and topological phases.
Very recently, Bismuth has been proposed as a material realization of a HOTI~\cite{schindler2018higher,schindler2018higher2}. 
While the material preserves time-reversal symmetry, the outlook is optimistic that eventual higher order phases with non-vanishing MQM will be discovered. Furthermore, our results provide a step toward an implementation to extract MQM in first-principles materials calculations. Finally, gapless phases like higher-order topological semimetals~\cite{lin2018topological} with nodes tilted in energy may also host a nontrivial MQM, the exploration of which we leave to future work.

{\bf Note added}: During the preparation of this manuscript we became aware of an independent work which also considers a similar method for calculating surface orbital magnetization~\cite{vanderbilt2022}.

\section*{Acknowledgements}
{We thank Roland Winkler, Penghao Zhu, and Atsuo Shitade for useful discussions. J.G., M.L., and T.L.H were supported by ARO MURI W911NF2020166. {J.G. also thanks NSF REU Grant No. PHY-1659598 for support.}}

\appendix

\begin{widetext}

\section{Multipole expansion for vector potential and boundary manifestation of magnetic multipoles}
\label{sec:Multipole expansion for vector potential and boundary manifestation of magnetic multipoles}

In this section, we derive an expression for the current distribution for various orders of magnetic multipoles in a continuous medium. The Einstein summation convention is assumed throughout, and we also assume that there is no bound charge or free current in the medium. 

\subsection{Some useful identities}
\label{sec:Some useful identities}

We start by discussing the assumptions and some identities that will be useful for later discussions. We consider a continuous medium with a boundary normal to ${\bf n}$. The medium is static, such that the continuity equation reduces to $\nabla\cdot {\mathbfcal J}=\partial_j\mathcal{J}_j=-\partial_t\rho=0$. This implies that $\int_{V}\nabla\cdot{\mathbfcal J}=\int_{\partial V}{\bf n}\cdot{\mathbfcal J}=0$ on the boundary by Gauss' theorem. In fact, we assume a stronger condition, namely
\begin{eqnarray}\begin{aligned}
\label{eq:BC_app}
{\bf n}\cdot{\mathbfcal J}=0,
\end{aligned}\end{eqnarray}
everywhere on the boundary. Armed with these conditions, we consider the following identities:
\begin{eqnarray}\begin{aligned}
\partial_j(r_i\mathcal{J}_j) &=\mathcal{J}_i\\
\partial_k(r_ir_j\mathcal{J}_k) &=2\mathcal{J}_ir_j+\epsilon_{ijk}({\bf r}\times{\mathbfcal J})_k\\
\partial_l(r_ir_jr_k\mathcal{J}_l) &=3\mathcal{J}_ir_jr_k+\epsilon_{ijl}r_k({\bf r}\times{\mathbfcal J})_l+\epsilon_{ikl}r_j({\bf r}\times{\mathbfcal J})_l.
\end{aligned}\end{eqnarray}
When integrated over the whole medium, the left hand sides of each of these equations all vanish for the boundary condition in Eq.~\eqref{eq:BC_app}, thus
\begin{eqnarray}\begin{aligned}
\label{eq:some_identities}
\int_V\mathcal{J}_i dV&=0\\
\int_V \mathcal{J}_ir_jdV &=-\frac{1}{2}\int_V\epsilon_{ijk}({\bf r}\times{\mathbfcal J})_kdV\\
\int_V \mathcal{J}_ir_jr_kdV &=-\frac{1}{3}\left(\epsilon_{ijl}\int_Vr_k({\bf r}\times{\mathbfcal J})_ldV+(j\leftrightarrow k)\right).
\end{aligned}\end{eqnarray}
We will make use of the identities above in the multipole expansion of the vector potential.

\subsection{The multipole expansion of the vector potential}

Let us consider a finite distribution of steady currents in vacuum. The vector potential at position ${\bf R}$ reads
\begin{eqnarray}\begin{aligned}
{\bf A}({\bf R}) = \frac{\mu_0}{4\pi}\int_V\frac{{\mathbfcal J}({\bf r})}{|{\bf R}-{\bf r}|}dV,
\end{aligned}\end{eqnarray}
where $dV=d^3r,$ and ${\mathbfcal J}({\bf r})$, is the current density at position ${\bf r}$. The factor $|{\bf R}-{\bf r}|^{-1}$ can be expanded in terms of Legendre polynomials. Explicitly, one can write
\begin{eqnarray}\begin{aligned}
|{\bf R}-{\bf r}|^{-1} = \frac{1}{R} + \frac{{\bf R}\cdot {\bf r}}{R^3} + \frac{3({\bf R}\cdot {\bf r})^2-R^2r^2}{2R^5}+...,
\end{aligned}\end{eqnarray}
such that the vector potential can be expanded in component form as
\begin{eqnarray}\begin{aligned}
A_i({\bf R}) = \frac{\mu_0}{4\pi}\left\{\frac{1}{R}\int_V \mathcal{J}_idV+\frac{R_j}{R^3}\int_V \mathcal{J}_ir_jdV+\frac{3R_jR_k-R^2\delta_{jk}}{2R^5}\int_V \mathcal{J}_ir_jr_kdV\right\}+...
\end{aligned}\end{eqnarray}

Analogously to the expansion of the scalar potential, the first term with $R^{-1}$ is interpreted as the magnetic monopole term (or magnetic charge), the second term with $R^{-2}$ is the magnetic dipole term, and the third is the magnetic quadrupole term. We are not interested in higher order terms so we will stop there. 
Now, the monopole term is \emph{always} zero from the first identity in Eq.~\eqref{eq:some_identities}. This essentially amounts to the fact that nonzero $\int_V \mathcal{J}_idV$ can arise only from a change in the medium's polarization, which we assume to be zero.
For the dipole term, if we define the magnetic dipole moment as
\begin{eqnarray}\begin{aligned}
{\bf M} = -\frac{1}{2}\int_V {\bf r}\times{\mathbfcal J}dV,
\end{aligned}\end{eqnarray}
then the second identity in Eq.~\eqref{eq:some_identities} gives a dipole term of
\begin{eqnarray}\begin{aligned}
R_j\int_V \mathcal{J}_ir_jdV = \epsilon_{ijk}R_jM_k = ({\bf R}\times{\bf M})_i.
\end{aligned}\end{eqnarray}
For the quadrupole, if we define
\begin{eqnarray}\begin{aligned}
M_{ij} = \frac{2}{3}\int_V r_i({\bf r}\times{\mathbfcal J})_jdV,
\end{aligned}\end{eqnarray}
then after using the third identity in Eq.~\eqref{eq:some_identities}, the quadrupole term in the expansion reads
\begin{eqnarray}\begin{aligned}
&(3R_jR_k-R^2\delta_{jk})\int_V \mathcal{J}_ir_jr_kdV \\
=&(3R_jR_k-R^2\delta_{jk})(\frac{-1}{3})(\epsilon_{ijl}\frac{3}{2}M_{kl}+\epsilon_{ikl}\frac{3}{2}M_{jl}) \\
=&-\frac{1}{2}(3R_jR_k-R^2\delta_{jk})(\epsilon_{ijl}M_{kl}+\epsilon_{ikl}M_{jl}) \\
=&-(3R_jR_k-R^2\delta_{jk})(\epsilon_{ijl}M_{kl}) \\
=&\epsilon_{ijl}(3R_kR_l-R^2\delta_{kl})M_{kj}.
\end{aligned}\end{eqnarray}

Writing the terms in this way, we find the desired expression for the vector potential in terms of magnetic multipole moments:
\begin{eqnarray*}\begin{aligned}
A_i({\bf R}) = \frac{\mu_0}{4\pi}\left\{\frac{({\bf R}\times{\bf M})_i}{R^3} + \frac{(3R_kR_l-R^2\delta_{kl})}{2R^5}\epsilon_{ijl}M_{kj}+...\right\},
\end{aligned}\end{eqnarray*}
 which is analogous to the multipole expansion for the scalar potential that gives rise to electric multipoles.

\subsection{Magnetic multipole moments in a macroscopic medium and their boundary manifestation}

We now consider multipole moments in a macroscopic medium. We divide the macroscopic material into unit cells, or microscopic voxels, over which the magnetic multipole densities are defined. 

The dipole contribution reads
\begin{eqnarray}\begin{aligned}
A^{(1)}_i({\bf R}) = \frac{\mu_0}{4\pi}\int_V dV \epsilon_{ikl}({\bf r}) \frac{\xi_k}{\xi^3}M_l,
\end{aligned}\end{eqnarray}
where
\begin{eqnarray}\begin{aligned}
\xi \equiv {\bf R} - {\bf r}.
\end{aligned}\end{eqnarray}
Now, because
\begin{eqnarray}\begin{aligned}
\frac{\xi_k}{\xi^3} = \frac{\partial}{\partial r_k}\frac{1}{\xi} \equiv \partial_k\frac{1}{\xi},
\end{aligned}\end{eqnarray}
we have
\begin{eqnarray}\begin{aligned}
A^{(1)}_i({\bf R}) &= \frac{\mu_0}{4\pi}\int_V dV \epsilon_{ikl}\left[\left(\partial_k\frac{1}{\xi}\right) M_l\right]\\
&= \frac{\mu_0}{4\pi}\int_{\partial V} dS \epsilon_{ikl}\left[n_k\frac{1}{\xi}M_l \right]-\frac{\mu_0}{4\pi}\int_V dV \epsilon_{ikl}\left[\partial_kM_l\frac{1}{\xi} \right],
\end{aligned}\end{eqnarray}
where $dS$ is a unit area element on the boundary.
Since both terms are of order $\xi^{-1}$, we can define the current densities as 
\begin{align}
    \mathcal{J}^\text{bulk}_i&=-\epsilon_{ikl}\partial_kM_l\label{eq:mdm_j_1}\\
\mathcal{J}^\text{boundary}_i&=\epsilon_{ikl}n_kM_l.\label{eq:mdm_j_2}
\end{align}

The quadrupole contribution reads
\begin{eqnarray}\begin{aligned}
A^{(2)}_i({\bf R}) = \frac{\mu_0}{4\pi}\int_V dV \frac{1}{2\xi^5}\epsilon_{ijl}(3\xi_k\xi_l-\xi^2\delta_{kl})M_{kj}({\bf r}).
\end{aligned}\end{eqnarray}
Now, because
\begin{eqnarray}\begin{aligned}
\partial_j\partial_i\frac{1}{\xi} = \frac{3\xi_i\xi_j-\xi^2\delta_{ij}}{\xi^5},
\end{aligned}\end{eqnarray}
we have
\begin{eqnarray}\begin{aligned}
A^{(2)}_i({\bf R}) = \frac{\mu_0}{4\pi}\int_V dV \frac{1}{2}\epsilon_{ijl}M_{kj}\partial_k\partial_l\frac{1}{\xi}.
\end{aligned}\end{eqnarray}
Again we integrate by parts, 
\begin{eqnarray}\begin{aligned}
M_{kj}\partial_k\partial_l\frac{1}{\xi} &=\partial_k(M_{kj}\partial_l\frac{1}{\xi})-(\partial_kM_{kj})\partial_l\frac{1}{\xi}\\
&=\partial_k\partial_l(M_{kj}\frac{1}{\xi})-2\partial_k\left[(\partial_lM_{kj})\frac{1}{\xi}\right]+(\partial_k\partial_lM_{kj})\frac{1}{\xi},
\end{aligned}\end{eqnarray}
such that
\begin{eqnarray}\begin{aligned}
A^{(2)}_i({\bf R}) &= \frac{\mu_0}{4\pi}\int_V dV \epsilon_{ijl}\left\{\frac{1}{2}\partial_k\partial_l(M_{kj}\frac{1}{\xi})-\partial_k\left[(\partial_lM_{kj})\frac{1}{\xi}\right]+\frac{1}{2}(\partial_k\partial_lM_{kj})\frac{1}{\xi}\right\} \\
&=\frac{\mu_0}{4\pi}\int_{\partial V} dS \epsilon_{ijl}\left\{\frac{1}{2}n_k\partial_l(M_{kj}\frac{1}{\xi})-n_k\left[(\partial_lM_{kj})\frac{1}{\xi}\right]\right\}+\frac{\mu_0}{4\pi}\int dV \epsilon_{ijl}\frac{1}{2}(\partial_k\partial_lM_{kj})\frac{1}{\xi} .\\
\end{aligned}\end{eqnarray}
Here we see that first and second terms can be expressed as boundary integrals. Generally, the boundary consists of different faces, which we label by the index $\alpha$, having discontinuous normal vectors ${\bf n}^\alpha$. We can then break up the first and second terms into sums of integrals on each face, such that 
\begin{eqnarray}\begin{aligned}
A^{(2)}_i({\bf R})  &=\frac{\mu_0}{4\pi}\sum_\alpha\int_{\partial V_\alpha} dS \epsilon_{ijl}\left\{\frac{1}{2}n^\alpha_k\partial_l(M_{kj}\frac{1}{\xi})-n^\alpha_k\left[(\partial_lM_{kj})\frac{1}{\xi}\right]\right\}+\frac{\mu_0}{4\pi}\int_V dV \epsilon_{ijl}\frac{1}{2}(\partial_k\partial_lM_{kj})\frac{1}{\xi} \\
&=\frac{\mu_0}{4\pi}\sum_{\alpha,\beta}\int_{\partial\partial V_{\alpha,\beta}} dL \epsilon_{ijl}\left\{\frac{1}{2}n^\alpha_kn^\beta_l(M_{kj}\frac{1}{\xi})\right\}-\frac{\mu_0}{4\pi}\sum_\alpha\int_{\partial V_\alpha} dS \epsilon_{ijl}\left\{n^\alpha_k\left[(\partial_lM_{kj})\frac{1}{\xi}\right]\right\}+\frac{\mu_0}{4\pi}\int_V dV \epsilon_{ijl}\frac{1}{2}(\partial_k\partial_lM_{kj})\frac{1}{\xi}, \\
\end{aligned}\end{eqnarray}
where $dL$ is a length element on hinges shared by adjacent surfaces. 
This gives the currents
\begin{align}
\mathcal{J}^\text{bulk}_i &=\frac{1}{2}\epsilon_{ijl}\partial_k\partial_lM_{jk} \label{eq:three_J_1}\\
\mathcal{J}^{\text{face},\alpha}_i &=-\epsilon_{ijl}n^\alpha_k\partial_lM_{jk}. \label{eq:three_J_2}\\
\mathcal{J}^{\text{hinge},\alpha,\beta}_i &=\frac{1}{2}\epsilon_{ikl}(n^{\alpha}_jn^{\beta}_lM_{jk}+n^{\beta}_jn^{\alpha}_lM_{jk}) \label{eq:three_J_3}
\end{align}
We note that the quantities in Eq.~\eqref{eq:three_J_1}-\eqref{eq:three_J_3} have different units. In particular, $\mathcal{J}^{\text{hinge},\alpha,\beta}_i$ has units of current, much like $M_{jk}$. Thus, we will denote the hinge current by ${J}^{\text{hinge},\alpha,\beta}_i$ to distinguish it from the bulk current density.
Furthermore, 
there are two terms in Eq.~\eqref{eq:three_J_3} because we treat $J^{\text{hinge},\alpha,\beta}_i$ and $J^{\text{hinge},\beta,\alpha}_i$ as same quantity, namely the hinge current flows on the hinge shared by the surfaces $\alpha$ and $\beta$. This is the hinge current in Eq.~\eqref{eq:Hinge_currents} in the main text.
We note that the surface current density $\mathcal{J}^{\text{face},\alpha}_i$ can be written as
\begin{align}
\label{eq:three_J_2_2}
    \mathcal{J}^{\text{face},\alpha}_i = \epsilon_{ilj}\partial_lK^\alpha_j
\end{align}
where $K^\alpha_j\equiv n^\alpha_kM_{jk}$. Eq.~\eqref{eq:three_J_2_2} resembles the one for the surface current in Eq.~\eqref{eq:mdm_j_1}. Hence we interpret $K^\alpha_j$ as a magnetization density (per unit area). 

\section{Condition for well-defined MQM}
\label{sec:Condition_zero_M}

\begin{figure}[ht!]
\centering
\includegraphics[width=0.5\columnwidth]{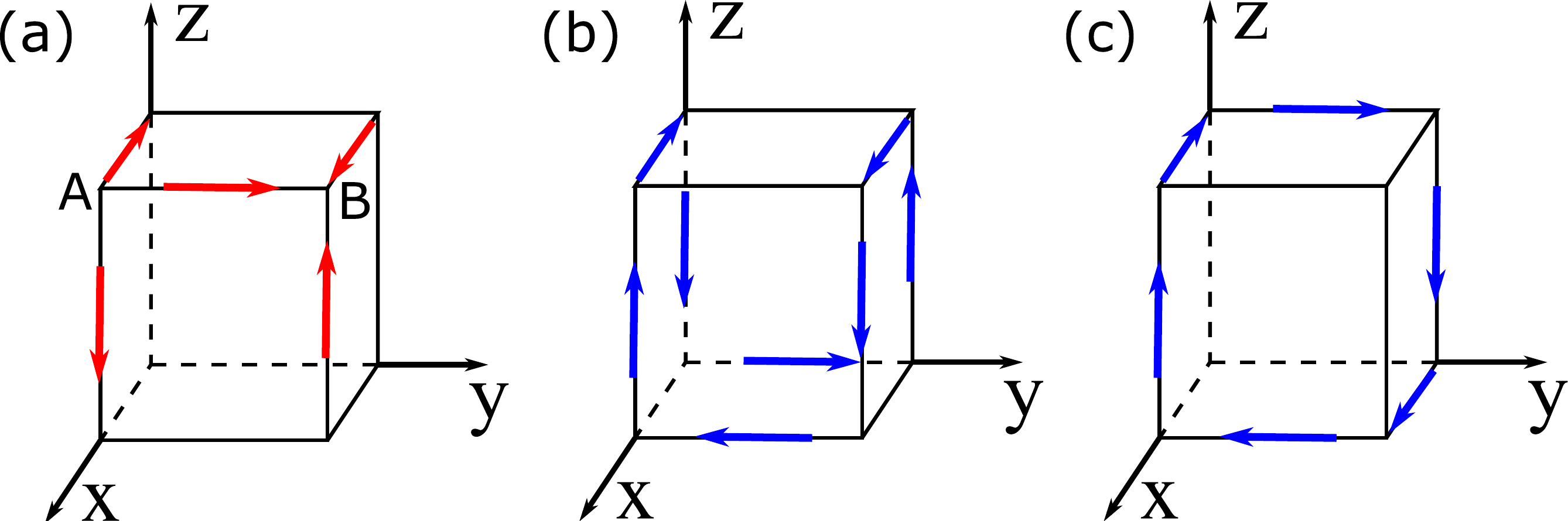}
\caption{(a) The hinge currents in Fig.~\ref{fig:1}(a) which contribute to $M_2$. Since one corner of the sample sits at the origin, the hinge currents that are not contributing have been omitted for simplicity. The red arrows indicate the positive directions of the currents according to Eq. ~\eqref{eq:Hinge_currents}. (b) A hinge current configuration with nonzero $M_3$. (c) A hinge current configuration with nonzero magnetization in all directions. For (b-c), the blue arrows indicate the direction in which the current flows. }
\label{fig:Other_HOTIs_v2}
\end{figure}

In this section, we prove that for the hinge configuration in Fig.~\ref{fig:1}(a), the system has zero magnetization if and only if the parallel hinges carry currents of the same magnitudes, namely $J_3^{+x,+y}=J_3^{-x,+y}=J_3^{+x,-y}=J_3^{-x,-y}$, and likewise for the other hinges. We limit our discussion to magnetization $M_2$ and assume that one corner of the sample sits at the origin for simplicity. Our conclusions do not change if another origin is chosen. As a result, we only show the hinge currents attached to the two vertices A and B in Fig.~\ref{fig:Other_HOTIs_v2}(a) because other hinge currents do not contribute to $M_2$. Since hinge currents are conserved at both vertices A and B, we have $J_2^{+z,+x}+J_3^{+x,-y}+J_1^{-y,+z}=0$ for A and $J_2^{+z,+x}+J_3^{+x,+y}+J_1^{+y,+z}=0$ for B. We thus have
\begin{eqnarray}\begin{aligned}
J_3^{+x,+y}+J_1^{+y,+z} = J_3^{+x,-y}+J_1^{-y,+z}.
\end{aligned}\end{eqnarray}
Moreover, based on the definition of magnetization in Eq.~\eqref{eq:def_MDM}, we have that
\begin{eqnarray}\begin{aligned}
M_2 = (J_3^{+x,+y}-J_1^{+y,+z}) - (J_3^{+x,-y}-J_1^{-y,+z}),
\end{aligned}\end{eqnarray}
where the contributions of $J_3^{-x,\pm y},J_1^{\pm y,-z}$ are zero for the chosen origin (hence not shown in Fig.~\ref{fig:Other_HOTIs_v2}(a)). If we set $M_2=0$ and combine these two equations, we have
\begin{eqnarray}\begin{aligned}
J_3^{+x,+y} - J_3^{+x,-y} = J_1^{+y,+z} - J_1^{-y,+z} = J_1^{-y,+z} - J_1^{+y,+z},
\end{aligned}\end{eqnarray}
leading to $J_1^{+y,+z} = J_1^{-y,+z}$ and $J_3^{+x,+y} = J_3^{+x,-y}$. In view of Fig.~\ref{fig:Other_HOTIs_v2}(a), this suggests that in order for $M_2$ to vanish, $J_1^{+y,+z}$ needs to have the same magnitude as $J_1^{-y,+z}$ but the opposite direction of propagation.
The same thing is true for $J_3^{+x,\pm y}$. We can generalize the argument to include other pairs of vertices, and in order to ensure ${\bf M}=0$ along all directions, the parallel hinges must carry currents of the identical magnitudes, and directions as indicated in Fig.~\ref{fig:1}(a).

We note that not all hinge current configurations are compatible with a well-defined MQM. As examples, in  Fig.~\ref{fig:Other_HOTIs_v2}(b-c), we illustrate two cases of hinge current configurations, with blue arrows showing the direction of actual current flow. For the system depicted in Fig.~\ref{fig:Other_HOTIs_v2}(b), the condition proved above is not respected, and hence the magnetization $M_3$ is nonzero. The situation is even worse for the system in Fig.~\ref{fig:Other_HOTIs_v2}(c), as there is nonzero magnetization in all directions. As a result, the MQM is not well-defined in these systems, and indeed certain HOTIs discussed in the literature exhibit these kinds of configurations~\cite{Pozo2019Quantization,zhu2020quantized}.
%

\end{widetext}

%

\end{document}